\documentclass{article}

\usepackage{microtype}
\usepackage{graphicx}
\usepackage{subfigure}
\usepackage{booktabs} 

\usepackage{hyperref}

\usepackage{amsmath,amssymb} 
\usepackage{mathtools}

\usepackage{algorithm}
\usepackage{algorithmic}
\usepackage{cases}

\usepackage[accepted]{icml2018}

\icmltitlerunning{Convergence Analysis and Design of Multi-block ADMM via Switched Control Theory}

\begin{document}

\twocolumn[
\icmltitle{Convergence Analysis and Design of Multi-block ADMM \\ via Switched Control Theory}




\begin{icmlauthorlist}
\icmlauthor{Jun Li}{to}
\icmlauthor{Hongfu Liu}{to}
\icmlauthor{Yue Wu}{to}
\icmlauthor{Yun Fu}{to}
\end{icmlauthorlist}

\icmlaffiliation{to}{Department of Electrical and Computer Engineering, Northeastern University, Boston, MA, USA.}

\icmlcorrespondingauthor{Jun Li}{junl.mldl@gmail.com}
\icmlcorrespondingauthor{Hongfu Liu}{liu.hongf@husky.neu.edu}
\icmlcorrespondingauthor{Yue Wu}{yuewu@ece.neu.edu}
\icmlcorrespondingauthor{Yun Fu}{yunfu@ece.neu.edu}

\icmlkeywords{Machine Learning, ICML}

\vskip 0.3in
]



\printAffiliationsAndNotice{\icmlEqualContribution} 

\begin{abstract}
We consider three challenges in multi-block Alternating Direction Method of Multipliers (\mbox{ADMM}): building convergence conditions for ADMM with any block (variable) sequence, finding available block sequences to be fit for \mbox{ADMM}, and designing useful parameter controllers for \mbox{ADMM} with unfixed parameters. To address these challenges, we develop a switched control framework for studying multi-block ADMM. First, since ADMM recursively and alternately updates the block-variables, it is converted into a discrete-time switched dynamical system. Second, we study exponential stability and stabilizability of the switched system for linear convergence analysis and design of ADMM by employing switched Lyapunov functions. Moreover, linear matrix inequalities conditions are proposed to ensure convergence of ADMM under arbitrary sequence, to find convergent sequences, and to design the fixed parameters. These conditions are checked and solved by employing semidefinite programming. 
\end{abstract}
\section{Introduction}
\label{introduction}
The alternating direction method of multipliers (ADMM) is usually applied to solve the following convex minimization problem with $N\geq3$ blocks of variables $\{\textbf{x}_i\}_{i=1}^{N}$:
\begin{align}
\begin{array}{l@{}l}
\min_{\textbf{x}_i\in\mathcal{X}_i,1\leq i\leq N}\ \sum_{i=1}^{N}f_i(\textbf{x}_i)\ \mbox{s.t.} \ \sum_{i=1}^{N} \textbf{A}_i\textbf{x}_i =\textbf{q},
\end{array}
\label{eq:mbcproblem}
\end{align}
where $\textbf{A}_i\in\mathbb{R}^{m\times n_i}$, $\textbf{q}\in\mathbb{R}^{m}$, $\mathcal{X}_i\subset\mathbb{R}^{n_i}$ are closed convex sets, and $f_i:\mathbb{R}^{n_i}\rightarrow \overline{\mathbb{R}}=\mathbb{R}\bigcup\{+\infty\}$ are closed proper convex functions. After the ADMM was originally proposed in the early 1970s \cite{Glowinski1975admm,Gabay1976ADM}, it has received increasing attention in wide applications, such as in machine learning, computer vision and signal processing \cite{Chang2016adadmm,Hong2016uaf,Wangjl2017dls}. Actually, the performance of these applications heavily depends on the convergence (at least acceptable accuracy) of ADMMs. It is well-known that the convergence of two-block ADMM has been proved in literature \cite{Eckstein1992admm}. When $N\geq 3$, however, the systematic convergence analysis of multi-block ADMM has been void for a long time.

Recently, global (linear) convergence and sublinear convergence of multi-block ADMM have been proved under the condition that functions are (strongly) convex and the penalty factor is restricted to a certain region \cite{Han2012admm,Lin2015mbadmm,Sun2015spADMM,Lin2015ADMMmv,Lin2016mbadmm,Li2016spADMM,xu2017laADMM,Lu2017mixadmm}. Moreover, \mbox{ADMMs} are also extended into parallel or distributed manners by using a Jacobi-type scheme \cite{Lu2016pl-admm,Dengw2017pmbadmm}. However, it is still unclear to systematically study multi-block ADMMs due to the following disadvantages. First, since these proofs are designed as an algorithm-by-algorithm basis \cite{Lessard2016iqc}, it lacks a unified framework to analyze the convergence problem of multi-block ADMM. Second, most of multi-block ADMM only consider a persistent block sequence (i.e., $\textbf{x}_1, \textbf{x}_2 \cdots, \textbf{x}_N$). This is conservative because it often leads to divergent algorithms. Third, ADMMs with unfixed parameters are usually divergent. In practice, it needs an effective method to design the fixed parameters. Forth, two-block ADMM is analyzed by adopting the integral quadratic constraint (IQC) framework \cite{Megretski1997iqc} from control theory, and formulating linear matrix equalities (LMIs) conditions for its linear convergence \cite{Nishihara2015gaadmm}. However, it only limits to the simple two-block ADMM. To study multi-block ADMM intensively, therefore, we focus on the following three basic challenges.


\textbf{Challenge 1:} \emph{Build sufficient conditions to guarantee that multi-block ADMM is convergent for any block sequence.} For the convergence problem, the first essential question is whether multi-block ADMM is convergent when there is no restriction on the block sequences. This basic challenge is called convergence under arbitrary (block) sequence.

\textbf{Challenge 2:} \emph{Find block sequences that makes multi-block ADMM convergent.} The arbitrary sequence is too strict to preserve the convergence of multi-block ADMM. However, it may be convergent under restricted block sequences. This leads us to find suitable sequences between the block-variables to ensure the convergence of multi-block ADMM, instead of the arbitrary sequence.

\textbf{Challenge 3:} \emph{Design fixed parameters that drive divergent multi-block ADMM to be convergent.} In fact, it is still possible to difficultly find a convergent block sequence in the challenge 2 due to the unsuitable parameters (e.g., penalty factors, damping factors, and matrices $\textbf{A}_i$ in Eq.~\eqref{eq:mbcproblem}). For example, a counter example is constructed to show the failure of three-block ADMM \cite{Chen2016ADMM}. This arouses the third basic challenge that how to construct fixed parameters to drive divergent ADMM to have a convergent trajectory.

In general, multi-block ADMM decomposes the convex problem \eqref{eq:mbcproblem} into $N$ smaller subproblems for the block-variables $\textbf{x}_i$. Specially, it alternately orchestrates switching between these subproblems, which are solved by recursively updating the corresponding block-variables. Clearly, this orchestration constitutes a block-variable switching sequence on the $N$ subproblems. Moreover, following the transformation of two-block ADMM \cite{Nishihara2015gaadmm}, each subproblem can be easily converted into a discrete-time dynamical subsystem. Thus, multi-block ADMM can be viewed as a \emph{discrete-time switched dynamical system} that consists of $N$ different subsystems and a switching rule between these subsystems \cite{LiberzonHespanhaMorseJun99a}. In particular, the related parameters are correspondingly changed into the weight matrices of the subsystems.

Based on above discussions, therefore, the three basic challenges of multi-block ADMM can be solved by analyzing the switched system with the switched control theory \cite{Lin2009sls}. \textbf{Challenge 1} is translated into building sufficient conditions for the stability of the switched system under arbitrary switching. Some LMIs conditions for the related parameters are constructed in subsection \ref{sec:problem1}. \textbf{Challenge 2} becomes to search for switching pathes on these conditions to ensure the stability of the switched system under suitable switching. A recursive search Algorithm \ref{alg:RecursiveSearch} is proposed in subsection \ref{sec:problem2}. \textbf{Challenge 3} turns to analyze the stabilizability of the switched system by designing feedback controller (regulative parameter matrices). Some controllers are developed in subsection \ref{sec:problem3}. This provides a practical approach to design the fixed parameters for multi-block ADMM instances. Overall, our main contributions are summarized as follows:
\begin{itemize}\vspace{-0.25cm}
\item To the best of our knowledge, we are the first to convert multi-block ADMM into a discrete-time switched dynamical system. The convergence of multi-block ADMM is analyzed and controlled by studying the stability and stabilizability of this switched system.
 \item \vspace{-0.15cm} We propose efficient LMIs conditions to ensure the linear convergence of multi-block ADMM under arbitrary sequence, and search for the convergent sequences on the block-variables. The size of LMIs scales with the number of the blocks.
 \item \vspace{-0.15cm} We design useful parameter controllers to drive multi-block ADMM to be convergent. These controllers are used to construct linear equality constraints for multi-block ADMM, and adjust the directions of trajectories of the variables in the feasible region. It is worth noting that the counter example \cite{Chen2016ADMM} can converge by designing a parameter controller.
\vspace{-0.2cm}\end{itemize}

\section{Preliminaries}
\label{sec:prenot}
In this section, we give some notations, and introduce the classical ADMMs and the discrete-time switched system.

\subsection{Notations}
We give some notations as follows. An index set $\mathcal{I}$ is denoted by $\{1,\cdots,N\hspace{-0.1cm}+\hspace{-0.1cm}1\}$. A $\textbf{P}$-norm of $\textbf{y}$, $\|\textbf{y}\|_{\textbf{P}}$, is denoted as $\sqrt{\textbf{y}^T\textbf{P}\textbf{y}}$. The condition number of $\textbf{A}$ is denoted as $\kappa_{\textbf{A}}=\sigma_1(\textbf{A})/\sigma_p(\textbf{A})$, where $\sigma_1(\textbf{A})$ and $\sigma_p(\textbf{A})$ denote the largest and smallest singular values of the matrix $\textbf{A}$. The Hadamard product of matrices $\textbf{A}$ and $\textbf{B}$ is denoted as $\textbf{A}\circ\textbf{B}$. We denote a $m\times m$ identity matrix and zero matrix by $\textbf{I}_m$ and $\textbf{0}_m$, respectively. A $m\times n$ all ones matrix is denoted as $\textbf{J}_{m\times n}$. The $i$-th $m\times m$ identity matrix with ones on the $i$-th main diagonal and zeros elsewhere is denoted as $\mathbb{I}_m^i=\mbox{diag}(0,\cdots,1,\cdots,0)$. We denote the Kronecker product of matrices $\textbf{A}$ and $\textbf{B}$ by $\textbf{A}\otimes\textbf{B}$. We denote a symmetric matrix $\left(\hspace{-0.2cm}
\begin{array}{cc}
\textbf{X} \hspace{-0.3cm}&  \textbf{Y} \\
\textbf{Y}^T \hspace{-0.3cm} & \textbf{Z} \\
\end{array}\hspace{-0.2cm}
\right)$ by $\left(\hspace{-0.2cm}
\begin{array}{cc}
\textbf{X} \hspace{-0.2cm} & \textbf{Y} \\
\star \hspace{-0.2cm} & \textbf{Z} \\
\end{array}\hspace{-0.2cm}
\right)$. A positive-definite matrix $\textbf{A}$ is denoted as $\textbf{A}>0$. In general, we let $\nabla f$ denote the gradient of $f$ if it is convex and differentiable, and let $\partial f$ denote the subdifferential of $f$ if it is convex. Suppose that the functions $f_i:\mathbb{R}^{n_i}\rightarrow \overline{\mathbb{R}}=\mathbb{R}\bigcup\{+\infty\}$ satisfy the following assumption.

\textbf{Assumption 1.} \emph{For any function $f_{ij}$, $i = 1, 2,\cdots, N$ and $j=1, 2,\cdots, n_i$, there exist constants $\nu_{ij}^{-}$ and $\nu_{ij}^{+}$  such that}
\begin{align}
\label{eq:fassupfuction}
\nu_{ij}^{-}\leq\frac{\mathfrak{f}_{ij}(x_{1})-\mathfrak{f}_{ij}(x_{2})}{x_{1}-x_{2}}\leq \nu_{ij}^{+},\ \forall x_{1}, x_{2}, x_{1}\neq x_{2},
\end{align}
\emph{where $\mathfrak{f}_{ij}(\cdot)$ is the gradient or subgradient of $f_{ij}(\cdot)$.}
For example, $\mathfrak{f}_{ij}(\cdot)\leftarrow\left\{
   \begin{array}{ll}
   =\nabla f_{ij}(\cdot), & \hbox{$f_{ij}$ is strong convex,}\\
   \in\partial f_{ij}(\cdot), & \hbox{$f_{ij}$ is convex.}
   \end{array}
   \right.$.

\subsection{Classical ADMMs}
To solve the problem \eqref{eq:mbcproblem}, we consider the augmented Lagrangian function:
\begin{equation}
\begin{array}{l@{}l}
&\mathcal{L}_{\beta}(\textbf{x}_1,\cdots,\textbf{x}_i,\cdots,\textbf{x}_N,\lambda)=\sum_{i=1}^{N}f_i(\textbf{x}_i)\\
&\ \ \ \ \ \ \ \ \ \ \ \ \ \ \ +\lambda^T (\sum_{i=1}^{N} \textbf{A}_i\textbf{x}_i-\textbf{q})+ \frac{\beta}{2}\|\sum_{i=1}^{N} \textbf{A}_i\textbf{x}_i -\textbf{q}\|_2^2,
\end{array} \nonumber
\end{equation} 
with the Lagrange multiplier $\lambda\in\mathbb{R}^{m}$ and a penalty parameter $\beta>0$. Roughly, most of ADMMs can be categorized into Gauss-Seidel ADMM and Jacobian ADMM. The Gauss-Seidel ADMMs update a variable $\textbf{x}_i$ by fixing others as their latest versions in a sequential manner \cite{Boyds2011dosl}, while the Jacobian ADMMs update all the $N$ block-variables $\textbf{x}_i$ $(1\leq i\leq N)$ in a parallel manner \cite{Dengw2017pmbadmm}. The iterative scheme of the Gauss-Seidel ADMM (GS-ADMM) is outlined below: for $1\leq i\leq N$,
\begin{align}
\left\{\hspace{-0.2cm}
\begin{array}{l@{}l}
\textbf{x}_i^{k+1}=&\mbox{arg}\min_{\textbf{x}_i} f_i(\textbf{x}_i)+\frac{\beta}{2}\|\sum_{j<i}\textbf{A}_j\textbf{x}_j^{k+1}\\
&  +\textbf{A}_i\textbf{x}_i+\sum_{j>i}\textbf{A}_j\textbf{x}_j^{k}+ \beta^{-1}\lambda^k-\textbf{q}\|_2^2\\
&  +\frac{1}{2}\|\textbf{x}_i-\textbf{x}_i^k\|_{\textbf{S}_i}^2,\\
\lambda^{k+1}=&\lambda^{k}-\gamma\beta(\sum_{j=1}^{N} \textbf{A}_j\textbf{x}_j^{k+1}-\textbf{q}),
\end{array}
\right.
\label{eq:gs-admmx}
\end{align}
and a general Jacobian ADMM, Proximal-Jacobian ADMM (PJ-ADMM) \cite{Dengw2017pmbadmm}, updates the variable $\textbf{x}_i$ in parallel by: for $1\leq i\leq N$,
\begin{align}
\left\{\hspace{-0.2cm}
\begin{array}{l@{}l}
\textbf{x}_i^{k+1}=&\mbox{arg}\min_{\textbf{x}_i} f_i(\textbf{x}_i)+\frac{\beta}{2}\|\sum_{j=1,\neq i}^{N}\textbf{A}_j\textbf{x}_j^{k}\\
&+\textbf{A}_i\textbf{x}_i+\beta^{-1}\lambda^k-\textbf{q}\|_2^2+\frac{1}{2}\|\textbf{x}_i-\textbf{x}_i^k\|_{\textbf{S}_i}^2,\\
\lambda^{k+1}=&\lambda^{k}-\gamma\beta(\sum_{j=1}^{N} \textbf{A}_j\textbf{x}_j^{k}-\textbf{q}),
\end{array}
\right.
\label{eq:pj-admmx}
\end{align}
where $\|\textbf{x}_i\|_{\textbf{S}_i}^2=\textbf{x}_i^T\textbf{S}_i\textbf{x}_i$, $\textbf{S}_i=\alpha_i\textbf{I}-\beta\textbf{A}_i^T\textbf{A}_i$ $(\alpha_i>0)$ and $\gamma>0$ is a damping parameter. When $\textbf{S}_i=0$, the ADMMs \eqref{eq:gs-admmx} and \eqref{eq:pj-admmx} can be rewritten as the standard ADMM \cite{Deng2016gadmm}. When $\textbf{S}_i=\alpha_i\textbf{I}$, it corresponds to the standard proximal method \cite{Parikh2014pa}. The convergent point of \eqref{eq:gs-admmx} and \eqref{eq:pj-admmx} is denoted as $\textbf{x}^\star=((\textbf{x}_1^\star)^T,\cdots,(\textbf{x}_N^\star)^T,(\lambda^\star)^T)^T$.

\subsection{Discrete-time Switched System}
Consider a discrete-time switched dynamical system
\begin{align}
\xi^{t+1}=\mathcal{B}_i\xi^{t}+\mathcal{C}_i \mathcal{G}(\xi^{t})+ \mathcal{D}_i\omega^{t}+\mathcal{E}_i\phi,\ \  i\in\mathcal{I},
\label{eq:dtsds}
\end{align}
with an initialized system state $\xi^0=\phi_{\xi}(0)$ and a control input $\omega^0=\phi_{\omega}(0)$, where $t\in\mathbb{Z}^+$, $\xi^{t}\in\mathbb{R}^{n_\xi}$ is the state, $\omega^{t}\in\mathbb{R}^{n_\omega}$ is the control input, $\phi$ is the offset input, $\mathcal{G}(\cdot)$ is a nonlinear function vector, $\mathcal{B}_i$ and $\mathcal{C}_i\in\mathbb{R}^{n_\xi\times n_\xi}$ are the state transition matrices, and $\mathcal{D}_i$ and $\mathcal{E}_i\in\mathbb{R}^{n_\xi\times n_\omega}$ are the control and offset input matrices. The system \eqref{eq:dtsds} is constructed for GS-ADMM, while $\mathcal{I}=\{1\}$, it becomes a discrete-time dynamical system for PJ-ADMM.

Usually, it is popular to study the stability and stabilizability of the system \eqref{eq:dtsds} by employing switched quadratic Lyapunov functions \cite{Daafouz2002slf}. A global Lyapunov function is constructed as $V(t,\xi^t)=(\xi^t)^T\textbf{P}_{\sigma(t)}\xi^t$, where $\sigma(t):\mathbb{Z}^+\rightarrow \mathcal{I}$. The index $\sigma(t)=i$ is called the active $i$-th subsystem at the discrete-time $t$, and $\textbf{P}_i$ $(i\in\mathcal{I})$ is a positive definite matrix for the Lyapunov function $V(t,\xi^t)$ of the $i$-th subsystem. The stability conditions of the system \eqref{eq:dtsds} is provided by proving that the Lyapunov function $V(t,\xi^t)$ is decreasing. For better developing the switched Lyapunov function for the system \eqref{eq:dtsds} in the section \ref{sec:SCT}, we introduce Finsler's Lemma \cite{Boyd1994lmi} as follow.

\textbf{Lemma 1} (Finsler's Lemma): \emph{Let $\textbf{y}\in\mathbb{R}^m$, $\textbf{P}\in\mathbb{R}^{n\times n} $ and $\textbf{H}\in\mathbb{R}^{n\times m}$ such that $rank(\textbf{H})=r <m$ and $\textbf{P}=\textbf{P}^T>0$. The following statements are equivalent:}

$\ \ \ $ \emph{1) $\textbf{y}^T\textbf{P}\textbf{y}<0$, $\forall\textbf{H}\textbf{y}=\textbf{0}, \textbf{y}\neq \textbf{0}$.}

$\ \ \ $ \emph{2) $\exists \textbf{Y}\in\mathbb{R}^{m\times n}: \textbf{P}+\textbf{YH}+\textbf{H}^T\textbf{Y}^T<0$.}

Finsler's Lemma has been previously used in the control literature \cite{Fang2004slf}. Lemma 1 aims to eliminate design variables in matrix inequalities. In addition, the definition of global exponential stability for the dynamical system \eqref{eq:dtsds} is now given.

\textbf{Definition 1.} \cite{Wu20101nd} \emph{The system \eqref{eq:dtsds} is said to be exponential stability, if there exist a constant $\chi>0$ and a factor $0<\tau<1$ such that}
\begin{align}
\|\xi^t-\xi^\star\|_2\leq \chi \tau^t\|\xi^0-\xi^\star\|_2.
\end{align}
Definition 1 shows that when the ADMMs are written as the system \eqref{eq:dtsds}, they will converge to $\xi^\star$ if the system \eqref{eq:dtsds} is exponential stable. The system state sequence $\{\xi^t\}$ is said to converge linearly if $0<\tau<1$. When $\tau$ (called convergence rate) is close to $0$, ADMMs are fast convergent, and vice versa.

\section{ADMMs as Switched Systems}
\label{sec:ADMM2system}
We convert the classical ADDMs into dynamical systems in this section. GS-ADMM and PJ-ADMM are transformed into discrete-time switched dynamical systems in subsections \ref{sec:gsADMM2system} and \ref{sec:pjADMM2system}, respectively. Before transforming ADMMs, we denote $\textbf{B}_i$, $\textbf{C}_i$, $\textbf{D}_i$ and $\textbf{E}_i$ as follow:
\begin{small}
\begin{align}
\label{eq:dtsdsmatrix1}
\textbf{B}_i=&\mathbb{I}_{N+1}^i\textbf{B},\ \textbf{C}_i=\mathbb{I}_{N+1}^i\textbf{C},\ \textbf{D}_i=\mathbb{I}_{N+1}^i\textbf{D},\ \textbf{E}_i=\mathbb{I}_{N+1}^i\textbf{E}, \\
\textbf{B}=&\left(\hspace{-0.2cm}
\begin{array}{ccccc}
1-\frac{\beta}{\widehat{\alpha}_1} \hspace{-0.2cm}& -\frac{\beta}{\widehat{\alpha}_1} \hspace{-0.2cm}&\cdots \hspace{-0.2cm} & -\frac{\beta}{\widehat{\alpha}_1} \hspace{-0.2cm}& -\frac{1}{\widehat{\alpha}_1} \\
-\frac{\beta}{\widehat{\alpha}_2} \hspace{-0.2cm}& 1-\frac{\beta}{\widehat{\alpha}_2} \hspace{-0.2cm}&\cdots \hspace{-0.2cm} & -\frac{\beta}{\widehat{\alpha}_2} \hspace{-0.2cm}& -\frac{1}{\widehat{\alpha}_2} \\
\vdots \hspace{-0.2cm}& \vdots \hspace{-0.2cm} & \ddots \hspace{-0.2cm} & \vdots \hspace{-0.2cm} & \vdots \\
-\frac{\beta}{\widehat{\alpha}_N} \hspace{-0.2cm}& -\frac{\beta}{\widehat{\alpha}_N} \hspace{-0.2cm} & \cdots \hspace{-0.2cm} & 1-\frac{\beta}{\widehat{\alpha}_N} \hspace{-0.2cm} & -\frac{1}{\widehat{\alpha}_N} \\
-\gamma\beta \hspace{-0.2cm}& -\gamma\beta \hspace{-0.2cm} & \cdots \hspace{-0.2cm} & -\gamma\beta \hspace{-0.2cm} & 0 \\
\end{array}\hspace{-0.2cm}
\right), \ \ \widehat{\alpha}_i=\frac{\alpha_i}{\|\textbf{A}_i\|^2}, \nonumber \\
\textbf{C}=&-\hbox{diag}\left(\frac{1}{\widehat{\alpha}_1},\cdots,\frac{1}{\widehat{\alpha}_i},\cdots\frac{1}{\widehat{\alpha}_N},0\right), \nonumber\\
\textbf{D}=&-\hbox{diag}\left(\frac{\beta}{\widehat{\alpha}_1}, \cdots,\frac{\beta}{\widehat{\alpha}_i}, \cdots,\frac{\beta}{\widehat{\alpha}_N},\gamma\beta\right),\  \textbf{E}=\textbf{D}. \nonumber
\end{align}
\end{small}We cast ADMMs as a discrete-time switched dynamical system with state sequences $\{\xi^t\}$, nonlinear functions $\mathcal{G}(\cdot)$, a control input $\omega^{t}$ and an offset input $\phi$. 

\emph{\textbf{State sequences $\{\xi^t\}$.}} Denote $\xi_i^k=\textbf{A}_i\textbf{x}_{i}^{k}$. We define sequences $\{\xi^t\in\mathbb{R}^{m(N\hspace{-0.05cm}+\hspace{-0.05cm}1)}\}$ for GS-ADMM in sequential manner by
\begin{align}
\xi^{t}=&\left((\xi_1^{k+1})^T,\cdots, (\xi_{i-1}^{k+1})^T, \right. \nonumber \\
\label{eq:xit}
      &\ \left.(\xi_{i}^{k})^T, (\xi_{i+1}^{k})^T, \cdots, (\xi_{N}^k)^T,(\lambda^k)^T \right)^T,
\end{align}
where $t=k(N\hspace{-0.1cm}+\hspace{-0.1cm}1)+i-1$ and $i$ is the number of the updated variables in the $k$-th iteration of GS-ADMM. Similar to GS-ADMM, we also define sequences $\{\xi^k\in\mathbb{R}^{m(N\hspace{-0.05cm}+\hspace{-0.05cm}1)}\}$ for PJ-ADMM in parallel manner by
\begin{align}
\label{eq:xik}
\xi^{k}=\left((\xi_1^{k})^T,\cdots, (\xi_{i}^{k})^T, \cdots, (\xi_{N}^k)^T,(\lambda^k)^T \right)^T.
\end{align}
In this paper, $\xi^{k}$ is denoted as $\xi^{t}$ for consistency.

\emph{\textbf{Nonlinear functions $\mathcal{G}(\cdot)$.}} Due to $\xi_i^k=\textbf{A}_i\textbf{x}_{i}^{k}$, $g_i:\mathbb{R}^{m}\rightarrow \overline{\mathbb{R}}$, where
$g_i=\left\{
   \begin{array}{ll}
   f_i\circ \textbf{A}_i^{-1}, & \hbox{$f_i$ is strong convex,} \\
   f_i\circ \textbf{A}_i^{\dag}+\mathcal{S}_{\mathcal{X}_i}, & \hbox{$f_i$ is convex.}
   \end{array}
   \right.$, $\textbf{A}_i^{\dag}$ is any left inverse of $\textbf{A}_i$, and $\mathcal{S}_{\mathcal{X}_i}$ is the $\{0,\infty\}$-indicator function. Based on the Assumption 1, a nonlinear function vector $\mathcal{G}:\mathbb{R}^{d(N+1)}\rightarrow \overline{\mathbb{R}}$ is denoted by
\begin{align}
\label{eq:actfunction}
\mathcal{G}(\cdot)=\left(\mathfrak{g}_1(\cdot),\cdots, \mathfrak{g}_{N}(\cdot),\mathfrak{g}_{\lambda}(\cdot) \right)^T\otimes \textbf{J}_{d\times 1},
\end{align}
where $\mathfrak{g}_i: \mathbb{R}\rightarrow \overline{\mathbb{R}}$ is used to simply represent a set $\{\mathfrak{g}_{il}\}_{1\leq l\leq d}$, $\mathfrak{g}_{il}(\cdot)$ is the gradient or subgradient of $g_{il}(\cdot)$, and $\mathfrak{g}_{\lambda}(\cdot)=0$. Actually, $\mathfrak{g}_i(\cdot)$ satisfies the following form:
\begin{align}
\mu_{i}^{-}\leq\frac{\mathfrak{g}_{i}(x_{1})-\mathfrak{g}_{i}(x_{2})}{x_{1}-x_{2}}\leq \mu_{i}^{+},\ \ \  \forall x_{1}, x_{2}, x_{1}\neq x_{2}, \nonumber
\end{align}
where $\mu_{i}^{-}=\min_j\{\nu_{ij}^{-}/\sigma_1^2(\textbf{A}_i), \nu_{ij}^{-}/\sigma_p^2(\textbf{A}_i), 1\leq j\leq n_i\}$, $\mu_{i}^{+}=\min_j\{\nu_{ij}^{+}/\sigma_1^2(\textbf{A}_i), \nu_{ij}^{+}/\sigma_p^2(\textbf{A}_i), 1\leq j\leq n_i\}$, $\nu_{ij}^{-}$ and $\nu_{ij}^{+}$ are defined in \eqref{eq:fassupfuction}. For presentation convenience in the following, we denote
\begin{align}
\begin{array}{l@{}l}
\textbf{F}_1&=\hbox{diag}\left(\mu_{1}^{-}\mu_{1}^{+},\cdots,\mu_{N}^{-}\mu_{N}^{+},0\right), \\
\textbf{F}_2&=\hbox{diag}\left(\frac{\mu_{1}^{-}+\mu_{1}^{+}}{2},\cdots,\frac{\mu_{N}^{-}+\mu_{N}^{+}}{2},0\right).
\end{array} \label{eq:F12}
\end{align}
\emph{\textbf{The control input $\omega^{t}$.}} In this paper, we aim also at designing the parameter feedback controllers,
\begin{align}
\omega^{t}=\mathcal{K}_i\xi^{t},\ \  i\in\mathcal{I},
\label{eq:controlinput}
\end{align}
where $\mathcal{K}_i=\textbf{K}_i\otimes \textbf{I}_d$, $\textbf{K}_i=\mathbb{I}_{N+1}^i\textbf{K},$ $\textbf{K}\in\mathbb{R}^{(N\hspace{-0.05cm}+\hspace{-0.05cm}1)\times (N\hspace{-0.05cm}+\hspace{-0.05cm}1)}$ is the control matrix. In subsection \ref{sec:problem3}, our main objective is to obtain a suitable parameter matrix $\textbf{K}$ for the system \eqref{eq:dtsds}.

\emph{\textbf{The offset input $\phi$.}} We consider the offset input $\phi=\left(\textbf{q}^T,\textbf{q}^T,\cdots,\textbf{0}^T\right)^T\in\mathbb{R}^{d(\hspace{-0.05cm}N+\hspace{-0.05cm}1)\times 1}$.
\subsection{GS-ADMM \eqref{eq:gs-admmx} as a Switched System \eqref{eq:gsdtsds}}
\label{sec:gsADMM2system}
Based on above denoted variables and Assumption 1, GS-ADMM can be written as a discrete-time switched system in the following Proposition 1.

\textbf{Proposition 1.} \emph{Denote system states $\xi^t$ in \eqref{eq:xit} with activation function $\mathcal{G}(\cdot)$ in \eqref{eq:actfunction}. GS-ADMM \eqref{eq:gs-admmx} is converted into the following switched system}
\begin{align}
\label{eq:gsdtsds}
\xi^{t+1}=&(\widehat{\textbf{B}}_i\otimes\textbf{I}_d)\xi^{t}+(\textbf{C}_i\otimes\textbf{I}_d) \mathcal{G}(\xi^{t}) \nonumber\\
& +(\textbf{E}_i\otimes\textbf{I}_d)\phi,\ \  i\in\mathcal{I},
\end{align}
\emph{where $\widehat{\textbf{B}}_i=\textbf{B}_i$ or $\widehat{\textbf{B}}_i=\textbf{B}_i+\textbf{D}_i\textbf{K}_i$. $\textbf{B}_i$, $\textbf{C}_i$, $\textbf{D}_i$, $\textbf{E}_i$ and $\textbf{K}_i$ are defined in \eqref{eq:dtsdsmatrix1} and \eqref{eq:controlinput}.}

Proposition 1 shows that GS-ADMM \eqref{eq:gs-admmx} is converted into the switched system \eqref{eq:gsdtsds}. The variable $\textbf{x}_i^{k+1}$ is updated by converting it into the system state $\xi^{t+1}$ with the $i$-th recursive subsystem $(\widehat{\textbf{B}}_i,\textbf{C}_i,\textbf{D}_i,\textbf{E}_i)$ for $1\leq i\leq N$. Similarly, the Lagrange multiplier $\lambda^{k+1}$ is transformed into the $N\hspace{-0.1cm}+\hspace{-0.1cm}1$-th recursive subsystem $(\widehat{\textbf{B}}_{N\hspace{-0.05cm}+\hspace{-0.05cm}1},\textbf{C}_{N\hspace{-0.05cm}+\hspace{-0.05cm}1},\textbf{D}_{N\hspace{-0.05cm}+\hspace{-0.05cm}1},\textbf{E}_{N\hspace{-0.05cm}+\hspace{-0.05cm}1})$. The arbitrarily updating sequences among the block variables correspond to the arbitrary switching between these subsystems on $\mathcal{I}\setminus\{N\hspace{-0.1cm}+\hspace{-0.1cm}1\}$. After computing all $\textbf{x}_i^{k+1}$, the system \eqref{eq:gsdtsds} switches to the $N\hspace{-0.1cm}+\hspace{-0.1cm}1$-th subsystem. When $\widehat{\textbf{B}}_i=\textbf{B}_i$, an effective way will be provided to study the challenge 1 and 2 of GS-ADMM by analyzing the stability of the switched system in subsection \ref{sec:problem1} and \ref{sec:problem2}. By using the controller $\omega^{t}$ in \eqref{eq:controlinput}, it also provides a new control method to make GS-ADMM convergent by $\widehat{\textbf{B}}_i=\textbf{B}_i+\textbf{D}_i\textbf{K}_i$, where $\textbf{K}_i$ is our solved control matrix in subsection \ref{sec:problem3}.

By using the convergent point $\textbf{x}^\star$ to further simplify the switched system \eqref{eq:gsdtsds} in the proposition 1, the equilibrium point $\xi^\star=((\xi_1^\star)^T,\cdots,(\xi_N^\star)^T(\lambda^\star)^T)^T=((\textbf{A}_1\textbf{x}_1^\star)^T,$ $\cdots,(\textbf{A}_N\textbf{x}_N^\star)^T,(\lambda^\star)^T)^T$ of \eqref{eq:gsdtsds} is shifted to the origin by the transformation $\overline{\xi}^t=\xi^t-\xi^\star$ and $\overline{\mathcal{G}}(\overline{\xi}^t)=\mathcal{G}(\xi^t-\xi^\star)-\mathcal{G}(\xi^\star)$, which converts the system to the following form:
\begin{align}
\overline{\xi}^{t+1}=(\widehat{\textbf{B}}_i\otimes \textbf{I}_d)\overline{\xi}^t+(\textbf{C}_i\otimes \textbf{I}_d)\overline{\mathcal{G}}(\overline{\xi}^{t}),\ \ i\in\mathcal{I}, \hspace{-0.2cm}
\label{eq:transformgsdtsds}
\end{align}
where $\overline{\mathcal{G}}(\cdot)$ satisfies $\mathcal{G}(\cdot)$ in \eqref{eq:actfunction} due to the transformation.

\subsection{PJ-ADMM \eqref{eq:pj-admmx} as a Dynamical System \eqref{eq:pjdtsds}}
\label{sec:pjADMM2system}

Following the conversion of GS-ADMM, PJ-ADMM can be considered as a dynamical system in Proposition 2.

\textbf{Proposition 2.} \emph{Denote system states $\xi^t$ in \eqref{eq:xik} with activation function $\mathcal{G}(\cdot)$ in \eqref{eq:actfunction}. PJ-ADMM \eqref{eq:pj-admmx} is converted into the following switched system}
\begin{align}
\label{eq:pjdtsds}
\xi^{t+1}=&(\widehat{\textbf{B}}\otimes\textbf{I}_d)\xi^{t}+(\textbf{C}\otimes\textbf{I}_d) \mathcal{G}(\xi^{t})+(\textbf{E}\otimes\textbf{I}_d)\phi, \hspace{-0.2cm}
\end{align}
\emph{where $\widehat{\textbf{B}}=\textbf{B}$ or $\widehat{\textbf{B}}=\textbf{B}+\textbf{D}\textbf{K}$. $\textbf{B}$, $\textbf{C}$, $\textbf{D}$, $\textbf{E}$ and $\textbf{K}$ are defined in \eqref{eq:dtsdsmatrix1} and \eqref{eq:controlinput}.}

Proposition 2 shows that PJ-ADMM is transformed into a dynamical system \eqref{eq:pjdtsds} due to the parallel manner. Compared to the Proposition 1, \eqref{eq:pjdtsds} is similar to the switched system \eqref{eq:gsdtsds} with $\mathcal{I}=\{1\}$. When $\widehat{\textbf{B}}=\textbf{B}$, the challenge 1 and 2 of PJ-ADMM will be analyzed by studying the stability of the dynamical system in subsection \ref{sec:problem1} and \ref{sec:problem2}. When $\widehat{\textbf{B}}=\textbf{B}+\textbf{D}\textbf{K}$, the parameter matrix $\textbf{K}$ is designed for the convergence of PJ-ADMM in subsection \ref{sec:problem3}.

Similar to the system \eqref{eq:transformgsdtsds}, by using the transformation the system \eqref{eq:pjdtsds} is converted to the following form:
\begin{align}
\overline{\xi}^{t+1}=&(\widehat{\textbf{B}}\otimes \textbf{I}_d)\overline{\xi}^t+(\textbf{C}\otimes \textbf{I}_d)\overline{\mathcal{G}}(\overline{\xi}^{t}).
\label{eq:transformpjdtsds}
\end{align}
\textbf{Remark 1:} \emph{Similar to the works \cite{Nishihara2015gaadmm,Lessard2016iqc}, the dimension of the parameters matrices in our systems \eqref{eq:gsdtsds} and \eqref{eq:pjdtsds} depends on the number of the block variables, instead of the size of $\textbf{A}_i$. Moreover, these parameters only depend on the norm of the $\textbf{A}_i$, that is, $\widehat{\alpha_i}=\frac{\alpha_i}{\|\textbf{A}_i\|^2}$. }

\textbf{Remark 2:}  \emph{Our systems \eqref{eq:gsdtsds} and \eqref{eq:pjdtsds} are different from the dynamical systems in the related literature \cite{Nishihara2015gaadmm,Lessard2016iqc,Hu2017dtnam,Hu2017usso}. First, in these literature, the gradient or subgradient of $f_i(\cdot)$ is regarded as the control input $\omega^{t}$, while it is used as a part of our systems since it is the self-driving force in the ADMMs. Second, we can make the systems stable by designing the parameter controller $\omega^{t}$, which provides a new regularization method for the ADMMs.}

\section{Switched Control Theory for ADMMs}
\label{sec:SCT}
In above section, multi-block ADMM are casted as discrete-time switched systems \eqref{eq:transformgsdtsds} and \eqref{eq:transformpjdtsds}. In this section, we solve the three basic challenges of multi-block ADMM by employing the switched control theory. First, we study stabilities of the switched systems for linear convergence of multi-block ADMM under arbitrary sequence in subsection \ref{sec:problem1}. Second, we find convergent sequences for GS-ADMM by searching for stable switching subsystems in subsection \ref{sec:problem2}. Third, we design parameter controllers to stabilize the switched systems for driving multi-block ADMM with unfixed parameters to be convergent in subsection \ref{sec:problem3}. Finally, a geometric interpretation for the designed parameter controllers is presented in subsection \ref{sec:geometric}.

\subsection{Building Convergence Conditions for Challenge 1}
\label{sec:problem1}

In this subsection, we propose linear convergence conditions for GS-ADMM under arbitrary sequence by employing switched quadratic Lyapunov functions (SQLFs) \cite{Daafouz2002slf} to study the exponential stability of the switched system \eqref{eq:transformgsdtsds} with $\widehat{\textbf{B}}_i=\textbf{B}_i$.

\textbf{Theorem 1.} \emph{Suppose that Assumption 1 holds. Fixing $0<\tau<1$, if there exist $(N\hspace{-0.1cm}+\hspace{-0.1cm}1)\hspace{-0.1cm}\times\hspace{-0.1cm} (N\hspace{-0.1cm}+\hspace{-0.1cm}1)$ positive definite matrices $\textbf{P}_i=\textbf{P}_i^T>0$ $(i\in\mathcal{I})$ and nonnegative constants $\Gamma=\mbox{diag}(\iota_1,\cdots,\iota_{N\hspace{-0.05cm}+\hspace{-0.05cm}1})>0$ such that the linear matrix inequalities: $\forall (i,j)\in\mathcal{I}\times\mathcal{I}$,}
\begin{align}
\label{eq:theorem1condition}
\left(\hspace{-0.2cm}
\begin{array}{cc}
\textbf{B}_i^T\textbf{P}_j\textbf{B}_i-\tau^2\textbf{P}_i+\Gamma\textbf{F}_1 & \textbf{B}_i^T\textbf{P}_j\textbf{C}_i+\Gamma\textbf{F}_2\\
\star  & \textbf{C}_i^T\textbf{P}_j\textbf{C}_i+\Gamma \\
\end{array}\hspace{-0.2cm}
\right)< 0,
\end{align}
\emph{where $\textbf{F}_1$ and $\textbf{F}_2$ are defined in \eqref{eq:F12}, then GS-ADMM under arbitrary sequence is linear convergent, that is,}
\begin{align}
\|\xi^{t}-\xi^\star\|_2< \min_{(i,j)\in\mathcal{I}\times\mathcal{I}}\left\{\sqrt{\kappa_{\textbf{P}_{ij}}}\right\}\tau^k \|\xi^{0}-\xi^\star\|_2,
\label{eq:theorem1convergence}
\end{align} 
where $\textbf{P}_{ij}=\mbox{diag}(\textbf{P}_i,\textbf{P}_j)$.

By using the Finsler's Lemma, we obtain a new Theorem.

\textbf{Theorem 2.} \emph{Suppose that Assumption 1 holds. Fixing $0<\tau<1$, if there exist $(N\hspace{-0.1cm}+\hspace{-0.1cm}1)\hspace{-0.1cm}\times\hspace{-0.1cm}(N\hspace{-0.1cm}+\hspace{-0.1cm}1)$ positive definite matrices $\textbf{P}_i=\textbf{P}_i^T>0$, matrices $\textbf{U}_{1i}, \textbf{U}_{2i}, \textbf{U}_{3i}\in\mathbb{R}^{(N\hspace{-0.05cm}+\hspace{-0.05cm}1)\times (N\hspace{-0.05cm}+\hspace{-0.05cm}1)}$ $(i\in\mathcal{I})$ and nonnegative constants $\Gamma=\mbox{diag}(\iota_1,\cdots,\iota_{N\hspace{-0.05cm}+\hspace{-0.05cm}1})>0$ such that the linear matrix inequalities: $\forall (i,j)\in\mathcal{I}\times\mathcal{I}$,} \begin{align}
\label{eq:theorem2condition}
\left(\hspace{-0.2cm}
\begin{array}{ccc}
\Xi_{11} & \Xi_{12}& \textbf{B}_i^T\textbf{U}_{3i}^T-\textbf{U}_{1i}\\
\star & \Xi_{22}& \textbf{C}_i^T\textbf{U}_{3i}^T-\textbf{U}_{2i}\\
\star & \star & \textbf{P}_j-\textbf{U}_{3i}^T-\textbf{U}_{3i}\\
\end{array} \hspace{-0.2cm}
\right)< 0,\ \ where
\end{align}
\emph{\resizebox{\linewidth}{!}{%
$
\begin{bmatrix*}[r]
\Xi_{11}=\textbf{U}_{1i}\textbf{B}_i+\textbf{B}_i^T\textbf{U}_{1i}^T-\tau^2\textbf{P}_i+\Gamma\textbf{F}_1,\ \ \Xi_{12}= \textbf{B}_i^T \\ \textbf{U}_{2i}^T+ \textbf{U}_{1i}\textbf{C}_i +\Gamma\textbf{F}_2,\ \ \Xi_{22}=\textbf{C}_i^T\textbf{U}_{2i}^T+\textbf{U}_{2i}\textbf{C}_i+\Gamma,
\end{bmatrix*},
$} $\textbf{F}_1$ and $\textbf{F}_2$ are defined in \eqref{eq:F12}, then GS-ADMM under arbitrary sequence is linear convergent.}

Theorems 1 and 2 provide two sufficient LMIs \mbox{conditions} to guarantee the linear convergence of GS-ADMM under arbitrary sequence in \eqref{eq:gs-admmx} by ensuring the exponential stability of the switched system \eqref{eq:transformgsdtsds}. These conditions directly construct the relationship between the parameters (i.e., $\widehat{\alpha}_i$, $\beta$, $\gamma$, $\textbf{F}_1$, $\textbf{F}_2$) and the convergence rate $\tau$ by using SQLFs. Based on the fixed parameters and $\tau$, the conditions \eqref{eq:theorem1condition} or \eqref{eq:theorem2condition} with the variables $\textbf{P}_i, \textbf{U}_{1i}, \textbf{U}_{2i}, \textbf{U}_{3i}$ and $\Gamma$ are easily solved by LMIs toolbox \cite{Boyd1994lmi} in the Matlab. Moreover, the minimal rate $\tau$ can be found by performing a binary search over $\tau$ to satisfy the conditions. To further simplify these conditions, we employ common quadratic Lyapunov functions (CQLFs) \cite{Mason2004clf}, that is, $V(t,\xi^t)=(\xi^t)^T\textbf{P}\xi^t$ with $\textbf{P}=\textbf{P}^T>0$, for the system \eqref{eq:transformgsdtsds} and have the following Corollary 1.

\textbf{Corollary 1.}  \emph{Suppose that Assumption 1 holds. Fixing $0<\tau<1$, if there exist a $(N\hspace{-0.1cm}+\hspace{-0.1cm}1)\hspace{-0.1cm}\times\hspace{-0.1cm} (N\hspace{-0.1cm}+\hspace{-0.1cm}1)$ positive definite matrix $\textbf{P}=\textbf{P}^T\succ0$ and nonnegative constants $\Gamma=\mbox{diag}(\iota_1,\cdots,\iota_{N\hspace{-0.05cm}+\hspace{-0.05cm}1})>0$ such that the linear matrix inequalities: $\forall i\in\mathcal{I}$,}
\begin{align}
\left(\hspace{-0.2cm}
\begin{array}{cc}
\textbf{B}_i^T\textbf{P}\textbf{B}_i-\tau^2\textbf{P}+\textbf{C}\textbf{F}_1 &\textbf{B}_i^T\textbf{P}\textbf{C}_i+\Gamma\textbf{F}_2\\
\star  & \textbf{C}_i^T\textbf{P}\textbf{C}_i+\Gamma \\
\end{array}\hspace{-0.2cm}
\right)< 0, \nonumber
\end{align} %
\emph{where $\textbf{F}_1$ and $\textbf{F}_2$ are defined in \eqref{eq:F12}, then GS-ADMM under arbitrary sequence is linear convergent.}

\textbf{Remark 3:} \emph{Our conditions for multi-block GS-ADMM come from the switched control theory \cite{LiberzonHespanhaMorseJun99a}, while the general analysis of two-block GS-ADMM \cite{Nishihara2015gaadmm} is based on the IQC method \cite{Megretski1997iqc}. Although the analysis can be extended for multi-block GS-ADMM by using a constant penalty parameter, it is conservative as this parameter is used to fix all block variables. }

\textbf{Remark 4:}\emph{ When the gradient of nonlinear functions $\mathfrak{g}_i(x)$ can be (approximately) linearized as $\mathfrak{g}_i(x)=\mathrm{g}_ix$ $(1\leq i\leq N)$, where $\mathrm{g}_i$ is a linearized parameter. For example, the functions $f_i$ are quadratic form. The related theories are presented in Section \ref{sec:slinear6} in supplementary materials.} 

Note that since the system \eqref{eq:transformpjdtsds} only has one subsystem with $\widehat{\textbf{B}}=\textbf{B}$ and $\textbf{C}$, these LMIs conditions with $i=j=1$ in Theorems 1, 2, and Corollary 1 are also suitable for the linear convergence of PJ-ADMM under arbitrary sequence. 

\subsection{Finding Switching Sequences for Challenge 2}
\label{sec:problem2}
In the above subsection, we provide some sufficient LMIs conditions to guarantee that GS-ADMM is linear convergent under arbitrary or given sequences on $\mathcal{I}$. However, arbitrary sequences are over strict to GS-ADMM since the different block sequences will result in different (convergent or divergent) trajectories. For instance, GS-ADMM \eqref{eq:gs-admmx} is difficult to be guaranteed the convergence even if each block variable is convergent. Thus, it is very important to build convergent block sequences on $\mathcal{I}$. We provide a solution to find the convergent sequences in this subsection.

According to the Theorems 1 and 2, these sufficient conditions are provided to search for a convergent block sequence on $\mathcal{I}$, which correspond to the block variables in GS-ADMM \eqref{eq:gs-admmx}. This problem is to find a switching sequence $S=S[1,\cdots,N\hspace{-0.1cm}+\hspace{-0.1cm}1]$ on $\mathcal{I}\times\mathcal{I}$, which satisfies the condition $\Psi_{S[i]S[i+1]}<0\ (1\leq i<N\hspace{-0.1cm}+\hspace{-0.1cm}1)$ and $\Psi_{S[1]S[n]}<0$ in \eqref{eq:theorem1condition} or \eqref{eq:theorem2condition}. It is similar to the classical $n$ \textbf{Queens} problem. Thus, the convergent block sequences can be found by \textbf{Algorithm \ref{alg:RecursiveSearch}} \textbf{RecursiveSearch}$(S,1)$. This backtracking algorithm is effective for the challenge 2 although it is simple and old.

Note that Corollary 1 provides another sufficient condition to check whether there exists a positive symmetric matrix such that all subsystems are strictly complete. Thus, GS-ADMM under all the block-switching sequences is convergent if the condition \eqref{eq:theorem1condition} is satisfied. the block sequences are not significant for PJ-ADMM because they updates the block variables in parallel manner. Thus, we do not consider the sequences problem in PJ-ADMM.

\subsection{Designing Parameter Controllers for Challenge 3}
\label{sec:problem3}

Given the related parameters $\widehat{\alpha}_i$, $\beta$ and $\gamma$, it is still difficult to find the convergence conditions and build the convergent block sequences for multi-block ADMM. In this subsection we can control the parameters by designing controllers to stabilize the switched systems for the convergence of multi-block ADMM.
For simplification, given a $N\hspace{-0.1cm}\times \hspace{-0.1cm}M$ matrix $\textbf{W}$, the $(i,j)$-th entry and the $i$-th row vector are denoted as $w_{ij}$ and $\textbf{w}_i=(w_{i1},\cdots,w_{ij},\cdots,w_{iM})$, respectively.

\subsubsection{Controlling GS-ADMM \eqref{eq:gs-admmx}}
GS-ADMM \eqref{eq:gs-admmx} can be controlled by parameter matrices $\textbf{K}_i$ in $\omega^{t}$ \eqref{eq:controlinput}.  $\textcolor[rgb]{0.98,0.00,0.00}{\textbf{K}_i}$ (The red characters are used to show the parameter controllers in ADMMs) only has the $i$-th row nonzero vector $\textcolor[rgb]{0.98,0.00,0.00}{\textbf{k}_i}=(\textcolor[rgb]{0.98,0.00,0.00}{k_{i1}},\textcolor[rgb]{0.98,0.00,0.00}{\cdots},\textcolor[rgb]{0.98,0.00,0.00}{k_{ij}},
\textcolor[rgb]{0.98,0.00,0.00}{\cdots},\textcolor[rgb]{0.98,0.00,0.00}{k_{i(N\hspace{-0.05cm}+\hspace{-0.05cm}1)}})$, which is used to improve the $i$-th updating rule in GS-ADMM \eqref{eq:gs-admmx} as the following form:
\begin{align}
\left\{\hspace{-0.2cm}
\begin{array}{l@{}l}
\textbf{x}_i^{k+1}=&\mbox{arg}\min_{\textbf{x}_i} f_i(\textbf{x}_i)+\frac{\beta}{2}\|(1+\textcolor[rgb]{0.98,0.00,0.00}{k_{ii}})\textbf{A}_i\textbf{x}_i \\
& +\sum_{j<i}(1+\textcolor[rgb]{0.98,0.00,0.00}{k_{ij}})\textbf{A}_j\textbf{x}_j^{k+1}\\
& +\sum_{j>i}(1+\textcolor[rgb]{0.98,0.00,0.00}{k_{ij}})\textbf{A}_j\textbf{x}_j^{k}\\
& + \beta^{-1}(1+\textcolor[rgb]{0.98,0.00,0.00}{k_{i(N\hspace{-0.05cm}+\hspace{-0.05cm}1)}})\lambda^k-\textbf{q} -\textcolor[rgb]{0.98,0.00,0.00}{\widehat{\textbf{q}}_i}\|_2^2 \\
&+\frac{1}{2}\|\textbf{x}_i-\textbf{x}_i^k\|_{\textcolor[rgb]{0.98,0.00,0.00}{\textbf{S}_i}}^2,\\
\lambda^{k+1}=&(1+\textcolor[rgb]{0.98,0.00,0.00}{k_{(N\hspace{-0.05cm}+\hspace{-0.05cm}1)(N\hspace{-0.05cm}+\hspace{-0.05cm}1)}})\lambda^{k}\\
&  -\gamma\beta(\sum_{j=1}^{N} (1+\textcolor[rgb]{0.98,0.00,0.00}{k_{(N\hspace{-0.05cm}+\hspace{-0.05cm}1)j}})\textbf{A}_j\textbf{x}_j^{k+1} \\
& -\textbf{q} -\textcolor[rgb]{0.98,0.00,0.00}{\widehat{\textbf{q}}_i}),
\end{array}
\right.
\label{eq:controlgs-admmx}
\end{align}
where $\textcolor[rgb]{0.98,0.00,0.00}{\textbf{S}_i}=\alpha_i\textbf{I}-\beta(1+\textcolor[rgb]{0.98,0.00,0.00}{k_{ii}})\textbf{A}_i^T\textbf{A}_i$, and $\textcolor[rgb]{0.98,0.00,0.00}{\widehat{\textbf{q}}_i}=\sum_{j=1}^N\textcolor[rgb]{0.98,0.00,0.00}{k_{ij}}\textbf{A}_j\textbf{x}_j^\star+\textcolor[rgb]{0.98,0.00,0.00}{k_{i(N\hspace{-0.05cm}+\hspace{-0.05cm}1)}}\lambda^\star$.
The improved GS-ADMM \eqref{eq:controlgs-admmx} is constructed by adding equality constraints
\begin{align}
\begin{array}{l@{}l}
\sum_{j=1}^N\textcolor[rgb]{0.98,0.00,0.00}{k_{ij}}\textbf{A}_j\textbf{x}_j+\textcolor[rgb]{0.98,0.00,0.00}{k_{i(N\hspace{-0.05cm}+\hspace{-0.05cm}1)}}\lambda=\textcolor[rgb]{0.98,0.00,0.00}{\widehat{\textbf{q}}_i}, i\in \mathcal{I},
\end{array}
\label{eq:controlplane}
\end{align}
into the original GS-ADMM \eqref{eq:gs-admmx}. $\textbf{k}_i$ is used to change the direction of the trajectory $\overline{\xi}_i^{t}=\textbf{A}_i\textbf{x}_i^t$ into a convergent region to improve the linear convergence of GS-ADMM. More analyses are discussed in subsection \ref{sec:geometric}.

\begin{algorithm}[t]
\caption{\textbf{RecursiveSearch}$(S,r)$}
   \label{alg:RecursiveSearch}
\begin{algorithmic}
   \STATE {\bfseries if} $r=n+1 \&\& \Psi_{S[1]S[n]}\preceq 0$ in Eqs. \eqref{eq:theorem1condition} or \eqref{eq:theorem2condition} 
   \STATE \hspace{0.3cm} print $S$
   \STATE {\bfseries else}
   \STATE \hspace{0.3cm} {\bfseries for} $j\leftarrow 1$ to $n$
   \STATE \hspace{0.8cm} $legal \leftarrow True$
   \STATE \hspace{0.3cm} {\bfseries for} $i\leftarrow 1$ to $r-1$
   \STATE \hspace{0.8cm} {\bfseries if} $\ \Psi_{ij}\succ0$ in Eqs. \eqref{eq:theorem1condition} or \eqref{eq:theorem2condition} 
   \STATE \hspace{1.2cm} $legal \leftarrow False$
   \STATE \hspace{0.4cm}{\bfseries if} $\ \ legal$
   \STATE \hspace{0.8cm} $S[r]\leftarrow j$
   \STATE \hspace{0.8cm} \textbf{RecursiveSearch}$(S,r+1)$
\end{algorithmic}
\end{algorithm}

Now, we give the following theory to obtain the parameter control matrices $\textbf{K}_i$. According to the Proposition 1 and the transformation, the improved GS-ADMM \eqref{eq:controlgs-admmx} is changed into the switched system \eqref{eq:transformgsdtsds} with $\widehat{\textbf{B}}_i=\textbf{B}_i+\textbf{D}_i\textbf{K}_i$. Based on Theorem 2, $\textbf{K}_i$ is designed in the following Theorem 3. Denote the $(i,i)$-th entry of diagonal matrix $\textbf{D}_i$ by $d_i$.

\textbf{Theorem 3.} \emph{Suppose that Assumption 1 holds. Fixing $0<\tau<1$, if there exist $(N\hspace{-0.1cm}+\hspace{-0.1cm}1)\hspace{-0.1cm}\times\hspace{-0.1cm}(N\hspace{-0.1cm}+\hspace{-0.1cm}1)$ positive definite matrices $\textbf{P}_i=\textbf{P}_i^T>0$, matrices $\textbf{U}_{3i}, \textbf{V}=\mbox{diag}(v_1,\cdots,v_{N\hspace{-0.05cm}+\hspace{-0.05cm}1})\in\mathbb{R}^{(N\hspace{-0.05cm}+\hspace{-0.05cm}1)\times (N\hspace{-0.05cm}+\hspace{-0.05cm}1)}$ $(i\in\mathcal{I})$ and nonnegative constants $\Gamma=\mbox{diag}(\iota_1,\cdots,\iota_{N\hspace{-0.05cm}+\hspace{-0.05cm}1})>0$ such that the linear matrix inequalities: $\forall (i,j)\in\mathcal{I}\times\mathcal{I}$,} \begin{align}
\label{eq:theorem3condition-1}
\left(\hspace{-0.2cm}
\begin{array}{ccc}
-\tau^2\textbf{P}_i+\Gamma\textbf{F}_1 & \Gamma\textbf{F}_2 & \textbf{B}_i^T\textbf{U}_{3i}^T+\mathbb{I}_{N+1}^i\textbf{V}\\
\star & \Gamma & \textbf{C}_i^T\textbf{U}_{3i}^T\\
\star & \star & \textbf{P}_j-\textbf{U}_{3i}^T-\textbf{U}_{3i}\\
\end{array} \hspace{-0.2cm}
\right)< 0,
\end{align}
\emph{where $\textbf{F}_1$ and $\textbf{F}_2$ are defined in \eqref{eq:F12}, then $\textbf{K}_i=\frac{v_{i}}{d_{i}}(\textbf{U}_{3i}^{-1}\mathbb{I}_{N+1}^i)^T\ (i\in\mathcal{I})$ drive that GS-ADMM under arbitrary sequence is linear convergent.}

Theorem 3 provides an efficient method to design the parameter controllers $\textbf{K}_i$ $(i\in\mathcal{I})$ to make GS-ADMM linear convergent by stabilizing the switched system \eqref{eq:transformgsdtsds}. By collecting all $\textbf{K}_i$, the parameter matrix $\textbf{K}$ is established by stacking the $i$-th row vector of $\textbf{K}_i$. 
%

\begin{figure*}[t]
\vskip -0.0in
\begin{center}
\centerline{\includegraphics[width=1.98\columnwidth]{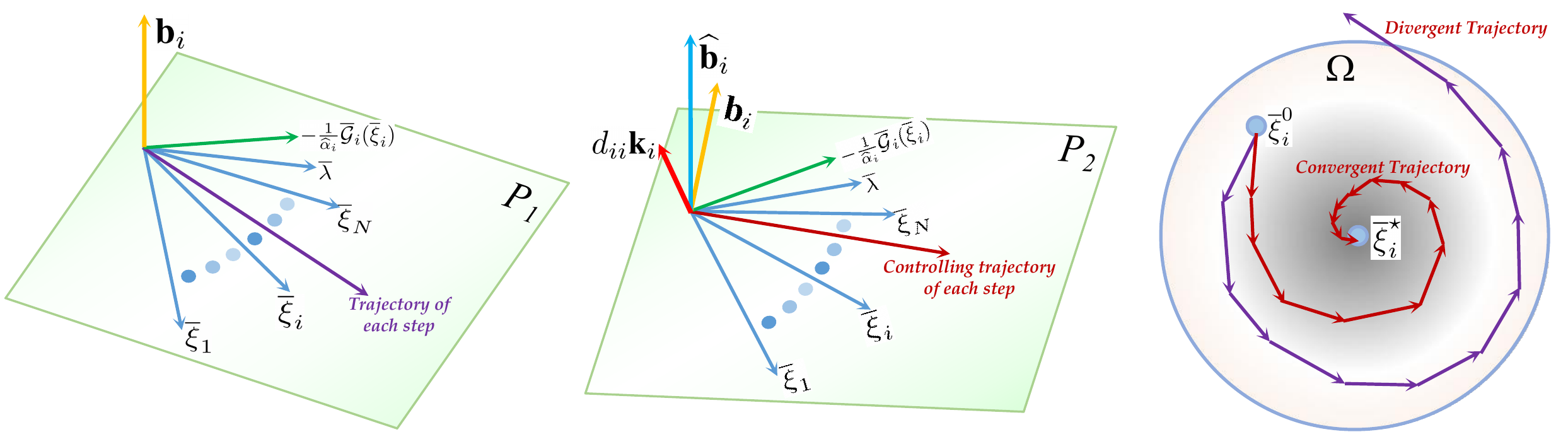}}
\vskip -0.15in
\caption{\emph{Left:} Based on a real space $(\overline{\xi}_1,\cdots,\overline{\xi}_N,\overline{\lambda})$, a purple trajectory of the $i$-th variable $\overline{\xi}_i$ in each updating step is synthetized by a hyperplane $P_1$ with the parameter normal vector $\textbf{b}_i$. \emph{Middle:} A scarlet trajectory of the $i$-th variable $\overline{\xi}_i$ is controlled by a hyperplane $P_2$ with a new parameter normal vector $\widehat{\textbf{b}}_i=\textbf{b}_i+k_{ii}\textbf{k}_i$. \emph{Right:} The divergent and convergent trajectories are plotted in the area $\Omega$. }
\label{fig:controller}
\end{center}
\vskip -0.25in
\end{figure*}

\subsubsection{Controlling PJ-ADMM \eqref{eq:pj-admmx}}
In this subsection, PJ-ADMM \eqref{eq:pj-admmx} can be controlled by the parameter matrix $\textbf{K}$. In particular, the $i$-th row nonzero vector $\textcolor[rgb]{0.98,0.00,0.00}{\textbf{k}_i}=(\textcolor[rgb]{0.98,0.00,0.00}{k_{i1}},\textcolor[rgb]{0.98,0.00,0.00}{\cdots},\textcolor[rgb]{0.98,0.00,0.00}{k_{ij}},
\textcolor[rgb]{0.98,0.00,0.00}{\cdots},\textcolor[rgb]{0.98,0.00,0.00}{k_{i(N\hspace{-0.05cm}+\hspace{-0.05cm}1)}})$ of  $\textcolor[rgb]{0.98,0.00,0.00}{\textbf{K}}$ is designed to revise the $i$-th updating rule in PJ-ADMM \eqref{eq:pj-admmx} as:
\begin{align}
\left\{\hspace{-0.2cm}
\begin{array}{l@{}l}
\textbf{x}_i^{k+1}=&\mbox{arg}\min_{\textbf{x}_i} f_i(\textbf{x}_i)+\frac{\beta}{2}\|(1+\textcolor[rgb]{0.98,0.00,0.00}{k_{ii}})\textbf{A}_i\textbf{x}_i \\
& +\sum_{j=1,\neq i}^{N}(1+\textcolor[rgb]{0.98,0.00,0.00}{k_{ij}})\textbf{A}_j\textbf{x}_j^{k}\\
& + \beta^{-1}(1+\textcolor[rgb]{0.98,0.00,0.00}{k_{i(N\hspace{-0.05cm}+\hspace{-0.05cm}1)}})\lambda^k-\textbf{q} -\textcolor[rgb]{0.98,0.00,0.00}{\widehat{\textbf{q}}_i}\|_2^2 \\
&+\frac{1}{2}\|\textbf{x}_i-\textbf{x}_i^k\|_{\textcolor[rgb]{0.98,0.00,0.00}{\textbf{S}_i}}^2,\\
\lambda^{k+1}=&(1+\textcolor[rgb]{0.98,0.00,0.00}{k_{(N\hspace{-0.05cm}+\hspace{-0.05cm}1)(N\hspace{-0.05cm}+\hspace{-0.05cm}1)}})\lambda^{k}\\
&  -\gamma\beta(\sum_{j=1}^{N} (1+\textcolor[rgb]{0.98,0.00,0.00}{k_{(N\hspace{-0.05cm}+\hspace{-0.05cm}1)j}})\textbf{A}_j\textbf{x}_j^{k} \\
& -\textbf{q} -\textcolor[rgb]{0.98,0.00,0.00}{\widehat{\textbf{q}}_i}),
\end{array}
\right.
\label{eq:controlpj-admmx}
\end{align}
where $\textcolor[rgb]{0.98,0.00,0.00}{\textbf{S}_i}$ and $\textcolor[rgb]{0.98,0.00,0.00}{\widehat{\textbf{q}}_i}$ are defined in \eqref{eq:controlgs-admmx}. Similar to the improved GS-ADMM, the improved PJ-ADMM \eqref{eq:controlpj-admmx} is constructed by adding the constraint \eqref{eq:controlplane} into the original GS-ADMM \eqref{eq:pj-admmx}. It is transformed into the switched system \eqref{eq:transformpjdtsds} with $\widehat{\textbf{B}}=\textbf{B}+\textbf{D}\textbf{K}$. Following the Theorem 3, $\textbf{K}$ is established in the following Corollary 2.

\textbf{Corollary 2.} \emph{Suppose that Assumption 1 holds. Fixing $0<\tau<1$, if there exist $(N\hspace{-0.1cm}+\hspace{-0.1cm}1)\hspace{-0.1cm}\times\hspace{-0.1cm}(N\hspace{-0.1cm}+\hspace{-0.1cm}1)$ positive definite matrices $\textbf{P}=\textbf{P}^T\succ0$, matrices $\textbf{U}_3, \textbf{V}, \textbf{H}\in\mathbb{R}^{(N\hspace{-0.05cm}+\hspace{-0.05cm}1)\times (N\hspace{-0.05cm}+\hspace{-0.05cm}1)}$ and nonnegative constants $\Gamma=\mbox{diag}(\iota_1,\cdots,\iota_{N\hspace{-0.05cm}+\hspace{-0.05cm}1})>0$ such that the linear matrix inequalities:}
\begin{align}
\label{eq:corollary2condition-1}
&\left(\hspace{-0.2cm}
\begin{array}{ccc}
-\tau^2\textbf{P}+\Gamma\textbf{F}_1 & \Gamma\textbf{F}_2 & \textbf{B}^T\textbf{U}_3^T+\textbf{H}^T\textbf{D}^T\\
\star & \Gamma & \textbf{C}^T\textbf{U}_3^T\\
\star & \star & \textbf{P}-\textbf{U}_3^T-\textbf{U}_3\\
\end{array} \hspace{-0.2cm}
\right)< 0, \\
\label{eq:corollary2condition-2}
& \ \ \ and \ \ \ \ \  \textbf{U}_3\textbf{D}=\textbf{D}\textbf{V},
\end{align}
\emph{where $\textbf{F}_1$ and $\textbf{F}_2$ are defined in \eqref{eq:F12}, then $\textbf{K}=\textbf{V}^{-1}\textbf{H}$ drives that PJ-ADMM is linear convergent.}

Corollary 2 builds the parameter controller $\textbf{K}$ to obtain the linear convergence of divergent PJ-ADMM by stabilizing the system \eqref{eq:transformpjdtsds}. In fact, the the trajectory of each variable is adjusted by the corresponding row vector of $\textbf{K}$. Following the linear transformation in Remark 5, the control theories are proposed in Section \ref{sec:slinear7} in supplementary materials. 

\textbf{Discussion:} The convergence of two-block ADMM \cite{Nishihara2015gaadmm} and some gradient methods (i.e. the Heavy-ball method and Nesterov's accelerated method) \cite{Lessard2016iqc,Hu2017dtnam} were proved by using the integral quadratic constraint (IQC) \cite{Megretski1997iqc}. In addition, IQC was also used to analyze the stochastic optimization methods \cite{Hu2017usso}, and an optimized network (OptNet) architecture was presented to integrate differentiable optimization problems (specifically, in the form of quadratic programs) \cite{Amos2017OptNet}. However, these methods do not analyze multi-block ADMM algorithms with multi-variable. More importantly, all of them do not consider the parameter control problem in these algorithms. We employ the switched control theory to design and control the multi-block ADMM algorithms.

\subsection{A Geometric Interpretation for Controllers}
\label{sec:geometric}
In this subsection, we give a geometric interpretation to better understand the parameter controllers $\textbf{K}_i$ for GS-ADMM and $\textbf{K}$ for PJ-ADMM.
We first show the geometric shapes of GS-ADMM and PJ-ADMM. Following the Propositions 1, 2 and the transformations in the subsection \ref{sec:ADMM2system}, GS-ADMM \eqref{eq:gs-admmx} and PJ-ADMM \eqref{eq:pj-admmx} are transformed into the system \eqref{eq:transformgsdtsds} with $\widehat{\textbf{B}}_i=\textbf{B}_i$ and the system \eqref{eq:transformpjdtsds} with $\widehat{\textbf{B}}=\textbf{B}$, respectively. Considering one block variable $\overline{\xi}_i=\xi_i-\xi_i^\star$, where $\xi_i=\textbf{A}_i\textbf{x}_i$ and $\xi_i^\star=\textbf{A}_i\textbf{x}_i^\star$, their updating rules are converted into the recursive forms:
\begin{small}
\begin{align}
\label{eq:simplegsdtds1}
&\begin{array}{l@{}l}
\overline{\xi}_i^{k+1}=&-\sum_{j<i}\frac{\beta}{\widehat{\alpha}_i}\overline{\xi}_j^{k+1}+(1-\frac{\beta}{\widehat{\alpha}_i})\overline{\xi}_i^k\\
&-\sum_{j>i}^N\frac{\beta}{\widehat{\alpha}_i}\overline{\xi}_j^{k}-\frac{1}{\widehat{\alpha}_i}\overline{\lambda}^k-\frac{1}{\widehat{\alpha}_i}\overline{\mathcal{G}}_i(\overline{\xi}_i^{k}),
\end{array} \\
\label{eq:simplepjdtds1}
&\begin{array}{l@{}l}
\overline{\xi}_i^{k+1}=&- \frac{1}{\widehat{\alpha}_i}\overline{\lambda}^k+(1-\frac{\beta}{\widehat{\alpha}_i})\overline{\xi}_i^k-\sum_{j=1,\neq i}^N\frac{\beta}{\widehat{\alpha}_i}\overline{\xi}_j^{k}\\
&-\frac{1}{\widehat{\alpha}_i}\overline{\mathcal{G}}_i(\overline{\xi}_i^{k}).
\end{array}
\end{align}
\end{small}Following in a real space $(\overline{\xi}_1,\cdots,\overline{\xi}_N,\overline{\lambda})$, the trajectory of the variable $\overline{\xi}_i$ is based on a gradient term $-\frac{1}{\widehat{\alpha}_i}\overline{\mathcal{G}}_i(\overline{\xi}_i)$ and a hyperplane $P_1$ pattern with a normal (or parameter) vector $\textbf{b}_i=[-\frac{\beta}{\widehat{\alpha}_1},\cdots,-\frac{\beta}{\widehat{\alpha}_{i-1}},1-\frac{\beta}{\widehat{\alpha}_i}, -\frac{\beta}{\widehat{\alpha}_{i+1}},\cdots,$ $-\frac{\beta}{\widehat{\alpha}_i}, -\frac{1}{\widehat{\alpha}_i}]$ in the left of Fig.~\ref{fig:controller}. Given an initial point $\overline{\xi}^{0}$, since the purple trajectory $\overline{\xi}_i^{t+1}$ is computed by \eqref{eq:simplegsdtds1} or \eqref{eq:simplepjdtds1}, its direction is determined by the parameter vector $\textbf{b}_i$, which corresponds to the $i$-th row vector of the parameter matrices $\textbf{B}_i$ or $\textbf{B}$.

When this purple trajectory is divergent in the right of Fig.~\ref{fig:controller}, it is desired to find a suitable normal vector to make $\overline{\xi}_i^{t+1}$ convergent. Our parameter controllers $\textbf{K}_i$ and $\textbf{K}$ are used to control GS-ADMM and PJ-ADMM. Their improved versions are shown in \eqref{eq:controlgs-admmx} and \eqref{eq:controlpj-admmx}, which are converted into the system \eqref{eq:transformgsdtsds} with $\widehat{\textbf{B}}_i=\textbf{B}_i+\textbf{D}_i\textbf{K}_i$  and the system \eqref{eq:transformpjdtsds} with $\widehat{\textbf{B}}=\textbf{B}+\textbf{D}\textbf{K}$, respectively. Similar to \eqref{eq:simplegsdtds1} and \eqref{eq:simplepjdtds1}, the new rules of $\overline{\xi}_i$ are changed into the recursive forms:
\begin{small}
\begin{align}
\label{eq:simplegsdtds2}
&\begin{array}{l@{}l}
\overline{\xi}_i^{k+1}=&-\sum_{j<i}\frac{\beta}{\widehat{\alpha}_i}(1+\textcolor[rgb]{0.98,0.00,0.00}{k_{ij}})\overline{\xi}_j^{k+1}\\
&+(1-\frac{\beta}{\widehat{\alpha}_i}(1+\textcolor[rgb]{0.98,0.00,0.00}{k_{ii}}))\overline{\xi}_i^k-\sum_{j>i}^N\frac{\beta}{\widehat{\alpha}_i}(1+\textcolor[rgb]{0.98,0.00,0.00}{k_{ij}})\overline{\xi}_j^{k}\\
&-\frac{1}{\widehat{\alpha}_i}(1+\textcolor[rgb]{0.98,0.00,0.00}{k_{i(N\hspace{-0.05cm}+\hspace{-0.05cm}1)}})\overline{\lambda}^k-\frac{1}{\widehat{\alpha}_i}\overline{\mathcal{G}}_i(\overline{\xi}_i^{k}),
\end{array} \\
\label{eq:simplepjdtds2}
&\begin{array}{l@{}l}
\overline{\xi}_i^{k+1}=&-\sum_{j=1,\neq i}^N\frac{\beta}{\widehat{\alpha}_i}(1+\textcolor[rgb]{0.98,0.00,0.00}{k_{ij}})\overline{\xi}_j^{k}\\
&+(1-\frac{\beta}{\widehat{\alpha}_i}(1+\textcolor[rgb]{0.98,0.00,0.00}{k_{ij}}))\overline{\xi}_i^k \\
& - \frac{1}{\widehat{\alpha}_i}(1+\textcolor[rgb]{0.98,0.00,0.00}{k_{i(N\hspace{-0.05cm}+\hspace{-0.05cm}1)}})\overline{\lambda}^k
-\frac{1}{\widehat{\alpha}_i}\overline{\mathcal{G}}_i(\overline{\xi}_i^{k}).
\end{array}
\end{align}
\end{small}The vector $\textbf{k}_i$ is designed to update $\textbf{b}_i$ as a new parameter vector $\widehat{\textbf{b}}_i$ for the hyperplane $P_2$ pattern in the middle of Fig.~\ref{fig:controller}, where $\widehat{\textbf{b}}_i=\textbf{b}_i+d_{ii}\textbf{k}_i$ and $d_{ii}=-\frac{\beta}{\widehat{\alpha}_i}$. Thus, the direction of the trajectory $\overline{\xi}_i^{t+1}$ is intuitively improved by the vector $\widehat{\textbf{b}}_i$. For example, the purple trajectory is revised as the scarlet trajectory in the right of Fig.~\ref{fig:controller}.


\textbf{Why the parameter controllers $\omega(\overline{\xi})=\textbf{D}\textbf{K}\overline{\xi}$ does not change the equilibrium point $\overline{\xi}^\star=0$ of PJ-ADMM?} First, the system \eqref{eq:transformpjdtsds} with $\widehat{\textbf{B}}=\textbf{B}$ has the unique equilibrium point $\overline{\xi}^\star=0$ since the problem \eqref{eq:mbcproblem} is convex. Second, $\overline{\xi}^\star=0$ is an equilibrium point of the system \eqref{eq:transformpjdtsds} with $\widehat{\textbf{B}}=\textbf{B}+\textbf{D}\textbf{K}$ because of $\omega(\overline{\xi}^\star)=\textbf{D}\textbf{K}\overline{\xi}^\star=0$. Third, the corresponding optimization problem is still convex because the feedback controller is a linear equality constraint. Thus, the equilibrium point $\overline{\xi}^\star=0$ is unique for the system \eqref{eq:transformpjdtsds} with $\widehat{\textbf{B}}=\textbf{B}+\textbf{D}\textbf{K}$. In fact, the key idea is to let the action of the ADMMs at any moment in time depend on the actual behavior of the the ADMMs that is being controlled. This idea imposes a certain 'smart' controller, which decides on the direction of the variables in the next moment. Similarly, $\omega(\overline{\xi}_i)=\textbf{D}_i\textbf{K}_i\overline{\xi}_i$ does not change the equilibrium point $\overline{\xi}_i^\star=0$ of GS-ADMM.

\textbf{An example.} Here we consider the following strongly convex problem with three variables \cite{Chen2016ADMM}:
\begin{align}
\label{eq:experiment3}
\min_{x_1,x_2,x_3}\ & 0.05x_1^2+0.05x_2^2+0.05x_3^2, \\
\emph{s.t.}\ & \textbf{A}_1x_1+\textbf{A}_2x_2+\textbf{A}_3x_3 =\textbf{0}, \nonumber
\end{align}
where $\textbf{A}_1=[1,1,1]^T$, $\textbf{A}_2=[1,1,2]^T$ and $\textbf{A}_3=[1, 2,2]^T$. The unique equilibrium point is $x_1=x_2=x_3=0$. By employing the PJ-ADMM with the parameters $\beta=1$, $\gamma=1$, and Lagrange multipliers $\lambda=[\lambda_1,\lambda_2,\lambda_3]^T$ to solve this problem \eqref{eq:experiment3}, it is divergent \cite{Chen2016ADMM}. 
Based on our Proposition 2, the problem \eqref{eq:experiment3} is transformed into a the dynamical system \eqref{eq:transformpjdtsds} with the state $\overline{\xi}=(x_1,x_2,x_3,\lambda_1,\lambda_2,\lambda_3)^T$. By employing the condition in Corollary C6 (in Section \ref{sec:slinear7}) or Corollary 2 with $\beta=1$, $\gamma=1$, $\alpha=1$ and $\tau=0.9$, we design a parameter matrix $\textbf{K}$ to control the divergent ADMM, where
\resizebox{\linewidth}{!}{%
$
\mathbf{K}=
\begin{bmatrix*}[r]
 -0.6500 &  0.0400 &  0.0100 &  0.1000 & -0.2000 & -0.0000 \\
  0.0400 & -0.7200 &  0.0200 & -0.0000 & -0.2000 &  0.2000 \\
  0.0100 &  0.0200 & -0.8800 &  0.1000 & -0.0000 &  0.1000 \\
  0.1000 & -0.0000 &  0.1000 & -1.0000 &  0.0000 &  0.0000 \\
 -0.2000 & -0.2000 &  0.0000 &  0.0000 & -1.0000 &  0.0000 \\
 -0.0000 &  0.2000 &  0.1000 & -0.0000 & -0.0000 & -1.0000 \\
\end{bmatrix*}.
$}
Each row vector of $\textbf{K}$ is used to construct a linear equality constraint for the corresponding variable. For instance, $\textbf{k}_1=(-0.65, 0.04, 0.01, 0.10, -0.20, 0.00)$ creates a constraint (hyperplane) $-0.65x_1+0.04x_2+0.01x_3+0.10\lambda_1+0.20\lambda_2+0\lambda_3=0$ for $x_1$. The normal vector $\textbf{k}_1$ to adjust the direction of the trajectory of $x_1$. Similar to \eqref{eq:controlpj-admmx}, the controller $\omega(\overline{\xi})=\textbf{K}\overline{\xi}$ is added into the the divergent ADMM to make it convergent for solving the problem \eqref{eq:experiment3}.

Note that all proofs of our theories and more numerical experiments are respectively provided in Sections \ref{sec:proofs} and \ref{sec:experiments} in supplementary materials.

\section{Conclusion}
\label{sec:con}
In this paper, we developed a switched Lyapunov framework to study the three basic problems of multi-block ADMM: building the convergence under arbitrary or given switching sequences, finding convergent switching sequences, and designing the fixed parameters. First, we employed switched quadratic Lyapunov functions to provide sufficient conditions to guarantee that multi-block ADMM is convergent. Second, based on the sufficient conditions, we proposed a backtracking algorithm to search for the convergent switching sequences. Third, we designed parameter controllers to make multi-block ADMM convergent. These controllers were in essence equality constraints to reduce the complexity of the original multi-block ADMM. Finally, a geometric interpretation showed the parameter controllers adjusted directions of trajectories of the variables. More importantly, this paper provided a new direction to analyze the multi-block ADMM by employing the switched control theory.


\bibliography{example_paper}
\bibliographystyle{icml2018}



\renewcommand\thefigure{S\arabic{figure}}
\renewcommand\thetable{S\arabic{table}}
\renewcommand\theequation{S\arabic{equation}}
\renewcommand\thesection{S\arabic{section}}



\twocolumn[
\icmltitle{Convergence Analysis and Design of ADMM via Switched \\
           Control Theory: Supplementary Material}

\section*{Classical ADMMs}
Classical ADMMs can be categorized into Gauss-Seidel ADMM and Jacobian ADMM. The iterative scheme of the Gauss-Seidel ADMM (GS-ADMM) is outlined below: for $1\leq i\leq N$,
\begin{align}
\left\{\hspace{-0.2cm}
\begin{array}{l@{}l}
\textbf{x}_i^{k+1}=&\mbox{arg}\min_{\textbf{x}_i} f_i(\textbf{x}_i)+\frac{\beta}{2}\|\sum_{j<i}\textbf{A}_j\textbf{x}_j^{k+1} +\textbf{A}_i\textbf{x}_i+\sum_{j>i}\textbf{A}_j\textbf{x}_j^{k}+ \beta^{-1}\lambda^k-\textbf{q}\|_2^2 +\frac{1}{2}\|\textbf{x}_i-\textbf{x}_i^k\|_{\textbf{S}_i}^2,\\
\lambda^{k+1}=&\lambda^{k}-\gamma\beta(\sum_{j=1}^{N} \textbf{A}_j\textbf{x}_j^{k+1}-\textbf{q}),
\end{array}
\right.
\label{eq:sgs-admmx}
\end{align}
A general Jacobian ADMM, Proximal-Jacobian ADMM (PJ-ADMM), updates the variable $\textbf{x}_i$ in parallel by: for $1\leq i\leq N$,
\begin{align}
\left\{\hspace{-0.2cm}
\begin{array}{l@{}l}
\textbf{x}_i^{k+1}=&\mbox{arg}\min_{\textbf{x}_i} f_i(\textbf{x}_i)+\frac{\beta}{2}\|\sum_{j=1,\neq i}^{N}\textbf{A}_j\textbf{x}_j^{k}+\textbf{A}_i\textbf{x}_i+\beta^{-1}\lambda^k-\textbf{q}\|_2^2+\frac{1}{2}\|\textbf{x}_i-\textbf{x}_i^k\|_{\textbf{S}_i}^2,\\
\lambda^{k+1}=&\lambda^{k}-\gamma\beta(\sum_{j=1}^{N} \textbf{A}_j\textbf{x}_j^{k}-\textbf{q}),
\end{array}
\right.
\label{eq:spj-admmx}
\end{align}
where $\|\textbf{x}_i\|_{\textbf{S}_i}^2=\textbf{x}_i^T\textbf{S}_i\textbf{x}_i$, $\textbf{S}_i=\alpha_i\textbf{I}-\beta\textbf{A}_i^T\textbf{A}_i$ $(\alpha_i>0)$ and $\gamma>0$ is a damping parameter.

\section{Building Convergence Conditions for Challenge 1 with Linear Transformation }
\label{sec:slinear6}
We consider a linear transformation of the switched systems. When the gradient of nonlinear functions $\mathfrak{g}_i(x)$ can be (approximately) linearized as $\mathfrak{g}_i(x)=\mathrm{g}_ix$ $(1\leq i\leq N)$, where $\mathrm{g}_i$ is a linearized parameter. For example, $f_i$ is a quadratic function, or the popular proximal function $\textbf{Prox}_\theta \mathfrak{g}(x)=\hbox{arg}\min_{y}\{\theta \mathfrak{g}(x)-||x-y||^2\}$ can be transformed as the following form $\textbf{Prox}_\theta \mathfrak{g}(x) = \mathrm{g}x$, where $\mathrm{g}=
\left\{
   \begin{array}{ll}
   \frac{|x|-\theta}{|x|}, & \hbox{if $|x|>\theta$,} \\
   0, & \hbox{otherwise.}
   \end{array}
   \right. $.
The nonlinear function $\mathcal{G}(\xi)$ in Eq. \eqref{eq:actfunction} in the main body is linearized as $\mathcal{G}(\xi)=(\textbf{G}_i\otimes\textbf{I}_d)\overline{\xi}$, where $\textbf{G}_i=\mathbb{I}_{N+1}^i\textbf{G}$ and $\textbf{G}=\hbox{diag}\left(\mathrm{g}_1,\cdots\mathrm{g}_N,0\right)$. Thus,

\subsection{linearized GS-ADMM}
The switched system in Eq. \eqref{eq:transformgsdtsds} in the main body for GS-ADMM is rewritten as a following linear switched system:
\begin{align}
\overline{\xi}^{t+1}=(\overline{\textbf{B}}_i\otimes \textbf{I}_d)\overline{\xi}^{t}, \ \ i\in\mathcal{I},
\label{eq:lineartransformgsdtsds}
\end{align}
where $\overline{\textbf{B}}_i=\widehat{\textbf{B}}_i+\textbf{C}_i\textbf{G}_i$. Based on the Corollary 1, Theorems 1 and 2 in the main body, we have the following Corollary C1.

\textbf{Corollary C1.} \emph{Suppose that Assumption 1 holds. Fixing $0<\tau<1$, the linearized GS-ADMM under arbitrary sequence is linear convergent, if there exist $(N\hspace{-0.1cm}+\hspace{-0.1cm}1)\hspace{-0.1cm}\times\hspace{-0.1cm}(N\hspace{-0.1cm}+\hspace{-0.1cm}1)$ positive definite matrices $\textbf{P}=\textbf{P}^T>0$,  $\textbf{P}_i=\textbf{P}_i^T>0$, and matrices $\textbf{U}_{1i}, \textbf{U}_{3i}\in\mathbb{R}^{(N\hspace{-0.05cm}+\hspace{-0.05cm}1)\times (N\hspace{-0.05cm}+\hspace{-0.05cm}1)}$ $(i\in\mathcal{I})$ such that the linear matrix inequalities:$\forall (i,j)\in\mathcal{I}\times\mathcal{I}$,}
\begin{align}
\label{eq:corollaryc1condition-1}
\left(\hspace{-0.2cm}
\begin{array}{cc}
\textbf{U}_{1i}\overline{\textbf{B}}_i+\overline{\textbf{B}}_i^T\textbf{U}_{1i}^T-\tau^2\textbf{P}_i &\overline{\textbf{B}}_i^T\textbf{U}_{3i}^T-\textbf{U}_{1i}\\
\star  & \textbf{P}_j-\textbf{U}_{3i}^T-\textbf{U}_{3i} \\
\end{array}\hspace{-0.2cm}
\right)< 0,
\end{align}
\emph{or $\forall (i,j)\in\mathcal{I}\times\mathcal{I}$,}
\begin{align}
\label{eq:corollaryc1condition-2}
\overline{\textbf{B}}_i^T\textbf{P}_j\overline{\textbf{B}}_i-\tau^2\textbf{P}_i < 0,
\end{align}
\emph{or $\forall i\in\mathcal{I}$,}
\begin{align}
\label{eq:corollaryc1condition-3}
\overline{\textbf{B}}_i^T\textbf{P}\overline{\textbf{B}}_i-\tau^2\textbf{P} < 0.
\end{align}
]

\twocolumn[
Two conditions in Corollary C1 are provided for the linear convergence of the linearized GS-ADMM as it is used to solve the convex problem in Eq. \eqref{eq:mbcproblem} in the main body with the linearized gradients of the functions, such as \emph{Quadratic Programming} problems.

\subsection{linearized PJ-ADMM}
Similar to the linearized GS-ADMM, the system in Eq. \eqref{eq:transformpjdtsds} for PJ-ADMM in the main body is rewritten as a following linear system:
\begin{align}
\overline{\xi}^{t+1}=(\overline{\textbf{B}}\otimes \textbf{I}_d)\overline{\xi}^{t},
\label{eq:lineartransformpjdtsds}
\end{align}
where $\overline{\textbf{B}}=\widehat{\textbf{B}}+\textbf{C}\textbf{G}$. Based on the Corollary C1, we have the following corollary C2.

\textbf{Corollary C2.} \emph{Suppose that Assumption 1 holds. Fixing $0<\tau<1$, the linearized PJ-ADMM is linear convergent, if there exist $(N\hspace{-0.1cm}+\hspace{-0.1cm}1)\hspace{-0.1cm}\times\hspace{-0.1cm}(N\hspace{-0.1cm}+\hspace{-0.1cm}1)$ positive definite matrices  $\textbf{P}=\textbf{P}^T>0$, and matrices $\textbf{U}_{1}, \textbf{U}_{3}\in\mathbb{R}^{(N\hspace{-0.05cm}+\hspace{-0.05cm}1)\times (N\hspace{-0.05cm}+\hspace{-0.05cm}1)}$ such that the linear matrix inequalities:}
\begin{align}
\label{eq:corollaryc2condition-1}
\left(\hspace{-0.2cm}
\begin{array}{cc}
\textbf{U}_{1}\overline{\textbf{B}}+\overline{\textbf{B}}^T\textbf{U}_{1}^T-\tau^2\textbf{P} &\overline{\textbf{B}}^T\textbf{U}_{3}^T-\textbf{U}_{1}\\
\star  & \textbf{P}-\textbf{U}_{3}^T-\textbf{U}_{3} \\
\end{array}\hspace{-0.2cm}
\right)< 0,
\end{align}
\emph{or }
\begin{align}
\label{eq:corollaryc2condition-2}
\overline{\textbf{B}}^T\textbf{P}\overline{\textbf{B}}-\tau^2\textbf{P} < 0.
\end{align}

\section{Designing Parameter Controllers for Challenge 3 with Linear Transformation }
\label{sec:slinear7}
Following the linear transformation in the above section, in this subsection we can control the parameters by designing controllers to stabilize the linear switched control systems for the convergence of ADMMs with linear transformation.

\subsection{Controlling GS-ADMM \eqref{eq:controllineartransformgsdtsds} with Linear Transformation}
The switched system in Eq. \eqref{eq:transformgsdtsds} in the main body is rewritten as the following linear switched control system:
\begin{align}
\overline{\xi}^{t+1}=((\overline{\textbf{B}}_i+\textbf{D}_i\textbf{K}_i)\otimes \textbf{I}_d)\overline{\xi}^{t}, \ \ i\in\mathcal{I},
\label{eq:controllineartransformgsdtsds}
\end{align}
where $\overline{\textbf{B}}_i=\textbf{B}_i+\textbf{C}_i\textbf{G}_i$ is defined in \eqref{eq:lineartransformgsdtsds}.

Based on the conditions \eqref{eq:corollaryc1condition-1} in the Corollary C1, it easily obtains the Corollary C3 to design the parameter controllers $\textbf{K}_i$ for the linear switched control system \eqref{eq:controllineartransformgsdtsds}.

\textbf{Corollary C3.} \emph{Suppose that Assumption 1 holds. Fixing $0<\tau<1$, if there exist $(N\hspace{-0.1cm}+\hspace{-0.1cm}1)\hspace{-0.1cm}\times\hspace{-0.1cm}(N\hspace{-0.1cm}+\hspace{-0.1cm}1)$ positive definite matrices $\textbf{P}_i=\textbf{P}_i^T\succ0$, any matrices $\textbf{U}_{3i}, \textbf{V}=\mbox{diag}(v_1,\cdots,v_{N\hspace{-0.05cm}+\hspace{-0.05cm}1})\in\mathbb{R}^{(N\hspace{-0.05cm}+\hspace{-0.05cm}1)\times (N\hspace{-0.05cm}+\hspace{-0.05cm}1)}$ $(i\in\mathcal{I})$ such that the linear matrix inequalities: $\forall (i,j)\in\mathcal{I}\times\mathcal{I}$,}
\begin{align}
\label{eq:Corollaryc3condition}
\left(\hspace{-0.2cm}
\begin{array}{ccc}
-\tau^2\textbf{P}_i & \overline{\textbf{B}}_i^T\textbf{U}_{3i}^T+\mathbb{I}_{N+1}^i\textbf{V}\\
\star & \textbf{P}_j-\textbf{U}_{3i}^T-\textbf{U}_{3i}\\
\end{array} \hspace{-0.2cm}
\right)\preceq 0,
\end{align}
\emph{then $\textbf{K}_i=\frac{v_i}{d_i}(\textbf{U}_{3i}^{-1}\mathbb{I}_{N+1}^i)^T\ (i\in\mathcal{I})$ drive that the linearized GS-ADMM under arbitrary sequence is linear convergent.}

According to the condition \eqref{eq:corollaryc1condition-2} in Corollary C1, it is difficult to solve the $\textbf{K}_i$ because one can easily show that the resulting inequalities are not jointly convex on $\textbf{P}_j$ and $\textbf{K}_i$. So, we consider the condition \eqref{eq:corollaryc1condition-3} in Corollary C1 by employing the CQLFs \cite{Mason2004clf} for the system \eqref{eq:controllineartransformgsdtsds}. Thus, the condition \eqref{eq:corollaryc1condition-3} is written as:
\begin{align}
(\overline{\textbf{B}}_i+\textbf{D}_i\textbf{K}_i)^T\textbf{P}(\overline{\textbf{B}}_i+\textbf{D}_i\textbf{K}_i)-\tau^2\textbf{P} < 0, \ \ \forall i\in\mathcal{I}. \nonumber
\end{align}
Generally, the problem of solving numerically the above form for $(\textbf{P}, \textbf{K}_i)$ is very difficult since it is nonconvex. In order to make this problem numerically well tractable, a sufficient condition is given in the following corollary by employing the Schur complement \cite{Boyd1994lmi}. Therefore, we have the following Corollary C4.

\textbf{Corollary C4.} \emph{Suppose that Assumption 1 holds. Fixing $0<\tau<1$, if there exist matrices $\textbf{N}_i$ $(i\in\mathcal{I})$ and positive definite matrices $\textbf{Q}=\textbf{Q}^T\succ0$ such that:}
\begin{align}
\label{eq:corollaryc4condition}
\left(\hspace{-0.2cm}
\begin{array}{cc}
-\textbf{Q} &\overline{\textbf{B}}_i\textbf{Q}+\textbf{D}_i\textbf{N}_i\\
\star & -\tau^2\textbf{Q} \\
\end{array}\hspace{-0.2cm}
\right)< 0, \ \ \forall i\in\mathcal{I},
\end{align}
\emph{then $\textbf{K}_i=\textbf{N}_i\textbf{Q}^{-1}\ (i\in\mathcal{I})$ drives that the linearized GS-ADMM under arbitrary sequence is linear convergent.}
]

\twocolumn[

\subsection{Controlling GS-ADMM \eqref{eq:controllineartransformpjdtsds} with Linear Transformation}
The switched system in Eq. (\eqref{eq:transformpjdtsds} in the main body is rewritten as the following linear switched control system:
\begin{align}
\overline{\xi}^{t+1}=((\overline{\textbf{B}}+\textbf{D}\textbf{K})\otimes \textbf{I}_d)\overline{\xi}^{t},
\label{eq:controllineartransformpjdtsds}
\end{align}
where $\overline{\textbf{B}}=\textbf{B}+\textbf{C}\textbf{G}$ is defined in \eqref{eq:lineartransformpjdtsds}.

Similar to the controlling GS-ADMM \eqref{eq:controllineartransformgsdtsds}, we have the following Corollary C5 based on the condition \eqref{eq:corollaryc2condition-1} in Corollary C2.

\textbf{Corollary C5.} \emph{Suppose that Assumption 1 holds. Fixing $0<\tau<1$, if there exist $(N\hspace{-0.1cm}+\hspace{-0.1cm}1)\hspace{-0.1cm}\times\hspace{-0.1cm}(N\hspace{-0.1cm}+\hspace{-0.1cm}1)$ positive definite matrices $\textbf{P}=\textbf{P}^T\succ0$, and  matrices $\textbf{U}, \textbf{V}, \textbf{H}\in\mathbb{R}^{(N\hspace{-0.05cm}+\hspace{-0.05cm}1)\times (N\hspace{-0.05cm}+\hspace{-0.05cm}1)}$ such that the linear matrix inequalities:}
\begin{align}
\label{eq:corollaryt5condition}
&\left(\hspace{-0.2cm}
\begin{array}{ccc}
-\tau^2\textbf{P}& \overline{\textbf{B}}^T\textbf{U}^T+\textbf{H}^T\textbf{D}^T\\
\star & \textbf{P}-\textbf{U}^T-\textbf{U}\\
\end{array} \hspace{-0.2cm}
\right)< 0, \ \ \ and \ \ \ \ \  \textbf{U}\textbf{D}=\textbf{D}\textbf{V},
\end{align}
\emph{then $\textbf{K}=\textbf{V}^{-1}\textbf{H}$ drives that the linearized PJ-ADMM is linear convergent.}

According to the condition \eqref{eq:corollaryc2condition-2} in Corollary C2, the following Corollary C6 is held by by employing the Schur complement \cite{Boyd1994lmi}.

\textbf{Corollary C6.} \emph{Suppose that Assumption 1 holds. Fixing $0<\tau<1$, if there exist matrices $\textbf{N}$ and positive definite matrices $\textbf{Q}=\textbf{Q}^T\succ0$ such that:}
\begin{align}
\label{eq:corollaryt6condition}
\left(\hspace{-0.2cm}
\begin{array}{cc}
-\textbf{Q} &\overline{\textbf{B}}\textbf{Q}+\textbf{D}\textbf{N}\\
\star & -\tau^2\textbf{Q} \\
\end{array}\hspace{-0.2cm}
\right)< 0, \ \ \forall i\in\mathcal{I},
\end{align}
\emph{then $\textbf{K}=\textbf{N}\textbf{Q}^{-1}$ drives that the linearized PJ-ADMM is linear convergent.}

\section{Proofs}
\label{sec:proofs}
\textbf{\emph{Proof of Proposition 1:}} It is proved by calculating the gradient or the subgradient of $\textbf{x}_i$-subproblem in GS-ADMM and set it to zero. Using the fact that $\textbf{A}_i$ has full rank and denoting that $\xi_i^k=\textbf{A}_i\textbf{x}_i^{k}$, the updating rule for $\textbf{x}_i$ from GS-ADMM \eqref{eq:sgs-admmx} can be rewritten as follows:
\begin{align}
\textbf{x}_i^{k+1}=&\textbf{A}_i^{-1}\mbox{arg}\min_{\xi_i} f_i(\textbf{A}_i^{-1}\xi_i)+\frac{\beta}{2}\|\sum_{j<i}\xi_j^{k+1}+\xi_i +\sum_{j>i}\xi_j^{k}+\beta^{-1}\lambda^k-\textbf{c}\|_2^2+\frac{\alpha_i}{2}\|\textbf{x}_i-\textbf{x}_i^k\|^2 -\frac{\beta}{2}\|\xi_i-\xi_i^k\|^2.
\label{eq:admmx1}
\end{align}
Since $\|\xi_i-\xi_i^k\|^2\leq\|\textbf{A}_i\|^2\|\textbf{x}_i-\textbf{x}_i^k\|^2$, multiplying through by $\textbf{A}_i$ the problem \eqref{eq:admmx1} is transformed into
\begin{align}
\xi_i^{k+1}=&\mbox{arg}\min_{\xi_i} g_i(\xi_i)+\frac{\beta}{2}\|\sum_{j<i}\xi_j^{k+1}+\xi_i+\sum_{j>i}\xi_j^{k}+\beta^{-1}\lambda^k-\textbf{c}\|_2^2 +\frac{\alpha_i}{2\|\textbf{A}_i\|^2}\|\xi_i-\xi_i^k\|^2-\frac{\beta}{2}\|\xi_i-\xi_i^k\|^2.
\label{eq:admmx2}
\end{align}
Denote $\widehat{\alpha}_i=\frac{\alpha_i}{\|\textbf{A}_i\|^2}$. The problem \eqref{eq:admmx2} implies that

\begin{align}
\xi_i^{k+1}=&\xi_i^{k}-\frac{\beta}{\widehat{\alpha}_i}(\sum_{j<i}\xi_j^{k+1}+\sum_{j\geq i}\xi_j^{k})-\frac{1}{\widehat{\alpha}_i}\lambda^k-\frac{1}{\widehat{\alpha}_i} \mathfrak{g}_i(\xi_i^{k+1})+\frac{\beta}{\widehat{\alpha}_i}\textbf{c}, \nonumber
\end{align}
where $\mathfrak{g}_i(\cdot)$ is defined in Eq. (10) in main body, and since $\mathfrak{g}_i(\textbf{x}_i^{k+1})$ is difficult to be directly calculated, it is approximated by $\mathfrak{g}_i(\textbf{x}_i^{k})$, and we have,
\begin{align}
\xi_i^{k+1}=&\xi_i^{k}-\frac{\beta}{\widehat{\alpha}_i}(\sum_{j<i}\xi_j^{k+1}+\sum_{j\geq i}\xi_j^{k})-\frac{1}{\widehat{\alpha}_i}\lambda^k-\frac{1}{\widehat{\alpha}_i} \mathfrak{g}_i(\xi_i^{k})+\frac{\beta}{\widehat{\alpha}_i}\textbf{c},
\label{eq:dsdadmmtdx}
\end{align}
We denote that
\begin{align}
\xi^{t}=&((\xi_1^{k+1})^T,\cdots,(\xi_{i-1}^{k+1})^T,(\xi_{i}^{k})^T,(\xi_{i+1}^{k})^T,\cdots,(\xi_{N}^k)^T,(\lambda^k)^T )^T, \nonumber \\
\xi^{t+1}=&((\xi_1^{k+1})^T,\cdots, (\xi_{i-1}^{k+1})^T,(\xi_{i}^{k+1})^T, (\xi_{i+1}^{k})^T, \cdots, (\xi_{N}^k)^T,(\lambda^k)^T)^T,\nonumber
\end{align}
where $t=k(N\hspace{-0.1cm}+\hspace{-0.1cm}1)+i-1$ and $i$ is the number of the updated variables in the $k$-th iteration of GS-ADMM.

]

\twocolumn[

 Thus, Eq. \eqref{eq:dsdadmmtdx} is transformed into the $i$-th subsystem
\begin{align}
\label{eq:sgsdtsds}
\xi^{t+1}=&(\widehat{\textbf{B}}_i\otimes\textbf{I}_d)\xi^{t}+(\textbf{C}_i\otimes\textbf{I}_d) \mathcal{G}(\xi^{t})+(\textbf{E}_i\otimes\textbf{I}_d)\phi,
\end{align}
where $\widehat{\textbf{B}}_i=\textbf{B}_i$, $\widehat{\textbf{B}}_i, \textbf{C}_i$ and $\textbf{E}_i$ are defined in Eq. \eqref{eq:dtsdsmatrix1} in main body.

By using the parameter feedback controllers $\omega^{t}=\mathcal{K}_i\xi^{t},\ i\in\mathcal{I}$,
where $\mathcal{K}_i=\textbf{K}_i\otimes \textbf{I}_d$, $\textbf{K}_i=\mathbb{I}_{N+1}^i\textbf{K}$ and $\textbf{K}\in\mathbb{R}^{(N\hspace{-0.05cm}+\hspace{-0.05cm}1)\times (N\hspace{-0.05cm}+\hspace{-0.05cm}1)}$ is the control matrix, it holds $\widehat{\textbf{B}}_i=\textbf{B}_i+\textbf{D}_i\textbf{K}_i$.

Similarly, we rewrite the updating rule for the Lagrange multiplier $\lambda$ in GS-ADMM \eqref{eq:sgs-admmx} as
\begin{align}
\label{eq:dsdadmmtdu}
\lambda^{k+1}=\lambda^{k}-\gamma\beta(\sum_{j=1}^{n}\xi_j^{k+1}-\textbf{c}),
\end{align}
which can be transformed into the $N\hspace{-0.1cm}+\hspace{-0.1cm}1$-th subsystem.

Together, all subsystems confirm the relationship matrices in Proposition 1. The proof is complete. $\Box$

$\\$
\textbf{\emph{Proof of Proposition 2:}} Similar to the proof of Proposition 1, Proposition 2 is easily proved and we omit it. $\Box$

$\\$
\emph{\textbf{Remark R1:}} When $\textbf{D}_i=\alpha_i\textbf{I}$ in GS-ADMM \eqref{eq:sgs-admmx} and PJ-ADMM \eqref{eq:spj-admmx}, the weight matrices in Proposition 1 and 2 are written as
\begin{align}
&\textbf{B}_i=\mathbb{I}_{N+1}^i\textbf{B},\ \textbf{C}_i=\mathbb{I}_{N+1}^i\textbf{C},\ \textbf{D}_i=\mathbb{I}_{N+1}^i\textbf{D},\ \textbf{E}_i=\mathbb{I}_{N+1}^i\textbf{E}, \\
&\textbf{B}=\left(\hspace{-0.2cm}
\begin{array}{ccccc}
\frac{\widehat{\alpha}_1}{\widehat{\alpha}_1+\beta} & -\frac{\beta}{\widehat{\alpha}_1+\beta} &  \cdots & -\frac{\beta}{\widehat{\alpha}_1+\beta} & -\frac{1}{\widehat{\alpha}_1+\beta} \\
-\frac{\beta}{\widehat{\alpha}_2+\beta} & \frac{\widehat{\alpha}_2}{\widehat{\alpha}_2+\beta} & \cdots & -\frac{\beta}{\widehat{\alpha}_2+\beta} & -\frac{1}{\widehat{\alpha}_2+\beta} \\
\vdots & \vdots & \ddots & \vdots & \vdots \\
-\frac{\beta}{\widehat{\alpha}_N+\beta} & -\frac{\beta}{\widehat{\alpha}_N+\beta} & \cdots  & \frac{\widehat{\alpha}_N}{\widehat{\alpha}_N+\beta} & -\frac{1}{\widehat{\alpha}_n+\beta} \\
-\gamma\beta & -\gamma\beta & \cdots & -\gamma\beta & 0 \\
\end{array}\hspace{-0.2cm}
\right), \ \ \widehat{\alpha}_i=\frac{\alpha_i}{\|\textbf{A}_i\|^2}, \\
& \textbf{C}=-\hbox{diag}\left(\frac{1}{\widehat{\alpha}_1+\beta},\cdots,\frac{1}{\widehat{\alpha}_i+\beta},\cdots\frac{1}{\widehat{\alpha}_n+\beta},0\right), \\
& \textbf{D}=-\hbox{diag}\left(\frac{\beta}{\widehat{\alpha}_1+\beta}, \cdots,\frac{\beta}{\widehat{\alpha}_i+\beta}, \cdots,\frac{\beta}{\widehat{\alpha}_n+\beta},\gamma\beta\right)^T, \textbf{E}=\textbf{D}.
\end{align}

$\\$
\textbf{\emph{Proof of Theorem 1:}} Construct a switched quadratic Lyapunov function \cite{Daafouz2002slf}:
\begin{align}
V(t,\xi^t)=(\overline{\xi}^t)^T\textbf{P}_{\sigma(t)}\otimes\textbf{I}_d\overline{\xi}^t, \nonumber
\end{align}
where $\textbf{P}_{\sigma(t)}\succ 0$ is to be determined. Defining

\begin{align}
\Delta V(t,\xi^t) =V(t+1,\xi^{t+1})-V(t,\xi^t) =(\overline{\xi}^{t+1})^T\textbf{P}_{\sigma(t+1)}\otimes\textbf{I}_d\overline{\xi}^{t+1}-(\overline{\xi}^t)^T\textbf{P}_{\sigma(t)}\otimes\textbf{I}_d\overline{\xi}^t, \nonumber
\end{align}
and adding a convergence rate $\tau$ into $\Delta V(t,\xi^t)$ yield
\begin{align}
\Delta V(t,\xi^t) \leq&(\overline{\xi}^{t+1})^T\textbf{P}_{\sigma(t+1)}\otimes\textbf{I}_d\overline{\xi}^{t+1} -\tau^2(\overline{\xi}^t)^T\textbf{P}_{\sigma(t)}\otimes\textbf{I}_d\overline{\xi}^t
\label{eq:deltav21}
\end{align}

From the assumption 1 and the nonlinear function $\mathcal{G}$ in Eq. \eqref{eq:actfunction} in the main body, it is easy to see that the following inequality holds for any nonnegative constants $\Gamma=\mbox{diag}(\iota_1,\cdots,\iota_{N\hspace{-0.05cm}+\hspace{-0.05cm}1})>0$,
\begin{align}
\label{eq:nonlinearfunctionineq}
\left(\hspace{-0.2cm}
\begin{array}{c}
\overline{\xi}^{t} \\
\overline{\mathcal{G}}(\overline{\xi}^{t})
\end{array}\hspace{-0.2cm}
\right)^T
\left(\hspace{-0.2cm}
\begin{array}{cc}
\Gamma\textbf{F}_1 & \Gamma\textbf{F}_2 \\
\star & \Gamma
\end{array}\hspace{-0.2cm}
\right)\otimes\textbf{I}_d
\left(\hspace{-0.2cm}
\begin{array}{c}
\overline{\xi}^{t} \\
\overline{\mathcal{G}}(\overline{\xi}^{t})
\end{array}\hspace{-0.2cm}
\right)\geq 0.
\end{align}

Adding the inequality \eqref{eq:nonlinearfunctionineq} and $\overline{\xi}^{k+1}$ in Eq. \eqref{eq:transformgsdtsds} in the main body to \eqref{eq:deltav21} yields
]

\twocolumn[
\begin{align}
\Delta V(t,\xi^t) \leq&((\textbf{B}_{\sigma(t)}\otimes \textbf{I}_d)\overline{\xi}^t+(\textbf{C}_{\sigma(t)}\otimes \textbf{I}_d)\overline{\mathcal{G}}(\overline{\xi}^{t}))^T\textbf{P}_{\sigma(t+1)}
((\textbf{B}_{\sigma(t)}\otimes \textbf{I}_d)\overline{\xi}^t+(\textbf{C}_{\sigma(t)}\otimes \textbf{I}_d)\overline{\mathcal{G}}(\overline{\xi}^{t}))\nonumber\\
&-\tau^2(\overline{\xi}^k)^T\textbf{P}_{\sigma(t)}\overline{\xi}^k+\left(\hspace{-0.2cm}
\begin{array}{c}
\overline{\xi}^{t} \\
\overline{\mathcal{G}}(\overline{\xi}^{t}) \\
\end{array}\hspace{-0.2cm}
\right)^T
\left(\hspace{-0.2cm}
\begin{array}{cc}
\Gamma\textbf{F}_1 & \Gamma\textbf{F}_2 \\
\star & \Gamma \\
\end{array}\hspace{-0.2cm}
\right)\otimes\textbf{I}_d
\left(\hspace{-0.2cm}
\begin{array}{c}
\overline{\xi}^{t} \\
\overline{\mathcal{G}}(\overline{\xi}^{t}) \\
\end{array}\hspace{-0.2cm}
\right) \nonumber \\
\label{eq:deltav23}
=&\left(\hspace{-0.2cm}
\begin{array}{c}
\overline{\xi}^{t} \\
\overline{\mathcal{G}}(\overline{\xi}^{t}) \\
\end{array}\hspace{-0.2cm}
\right)^T\Psi_{\sigma(t)\sigma(t+1)}\otimes\textbf{I}_d\left(\hspace{-0.2cm}
\begin{array}{c}
\overline{\xi}^{t} \\
\overline{\mathcal{G}}(\overline{\xi}^{t}) \\
\end{array}\hspace{-0.2cm}
\right),
\end{align}
where $\Psi_{\sigma(t)\sigma(t+1)}=\left(\hspace{-0.2cm}
\begin{array}{cc}
\textbf{B}_{\sigma(t)}^T\textbf{P}_{\sigma(t+1)}\textbf{B}_{\sigma(t)}-\tau^2\textbf{P}_{\sigma(t)}+\Gamma\textbf{F}_1 & \textbf{B}_{\sigma(t)}^T\textbf{P}_{\sigma(t+1)}\textbf{C}_{\sigma(t)}+\Gamma\textbf{F}_2\\
\star  & \textbf{C}_{\sigma(t)}^T\textbf{P}_{\sigma(t+1)}\textbf{C}_{\sigma(t)}+\Gamma \\
\end{array}\hspace{-0.2cm}
\right)$.

Since $\Delta V(t,\xi^t) $ will be satisfied under arbitrary switching sequences, it has to hold for the special configuration $\sigma(t)=i$, $\sigma(t+1)=j$, and for all $\overline{\xi}^t\in\mathbb{R}^{d(N\hspace{-0.05cm}+\hspace{-0.05cm}1)}$. Then if $\Psi_{ij}< 0$ for all $(i,j)\in\mathcal{I}\times\mathcal{I}$, then $\Delta V(t,\xi^t)< 0$ for any $\overline{\xi}^t\neq\textbf{0}$, that is, the switched system under arbitrary switching sequences is stable.

Next, we prove that the switched system is exponentially stable. By using the transformation $\overline{\xi}^t=\xi^t-\xi^\star$ in \eqref{eq:deltav21}, and the nonnegativity of \eqref{eq:deltav21}, we have
\begin{align}
(\xi^{l}-\xi^\star)^T\textbf{P}_{\sigma(l)}(\xi^{l}-\xi^\star)\leq
\tau^2(\xi^{l-1}-\xi^\star)^T\textbf{P}_{\sigma(l-1)}(\xi^{l-1}-\xi^\star). \nonumber
\end{align}
Inducting from $l = 1$ to $ t$, we see that for all $t$
\begin{align}
(\xi^{t}-\xi^\star)^T\textbf{P}_{\sigma(l)}(\xi^{t}-\xi^\star)\leq
\tau^{2t}(\xi^{0}-\xi^\star)^T\textbf{P}_{\sigma(0)}(\xi^{0}-\xi^\star). \nonumber
\end{align}
Thus, it holds Eq. \eqref{eq:theorem1convergence} in the main body. The proof is complete. $\Box$

$\\$
\textbf{\emph{Proof of Theorem 2:}} Recall that the requirement $\Delta V(t,\xi^t)<0$ in \eqref{eq:deltav21} by adding the inequality \eqref{eq:nonlinearfunctionineq}, $\forall\xi^t\neq 0$ can be stated as $\exists\textbf{P}_i=\textbf{P}_i^T\succ0,\textbf{P}_j=\textbf{P}_j^T\succ0$ such that
\begin{align}
\textbf{y}^T\textbf{P}\otimes\textbf{I}_d
\textbf{y}<0, \ \ \
\forall\ \textbf{H}\textbf{y}=0, \ \ \
\textbf{y}\neq 0,
\end{align}
where
\begin{align}
\textbf{y}=\left(\hspace{-0.2cm}
\begin{array}{c}
\overline{\xi}^{t} \\
\overline{\mathcal{G}}(\overline{\xi}^{t}) \\
\overline{\xi}^{t+1}
\end{array}\hspace{-0.2cm}
\right),\ \ \
\textbf{P}=\left(\hspace{-0.2cm}
\begin{array}{ccc}
-\tau^2\textbf{P}_i+\Gamma\textbf{F}_1 & \Gamma\textbf{F}_2 & 0 \\
\star & \Gamma & 0\\
\star & \star  & \textbf{P}_j\\
\end{array}\hspace{-0.2cm}
\right),\ \ \
\textbf{H}=\left(\hspace{-0.2cm}
\begin{array}{ccc}
\textbf{B}_i \hspace{-0.2cm}& \textbf{C}_i \hspace{-0.2cm}& -\textbf{I}
\end{array}\hspace{-0.2cm}
\right).
\end{align}

Applying the Lemma 1 in the main body with $\textbf{y}$, $\textbf{P}$, $\textbf{H}$, and
$\textbf{Y}=\left(\hspace{-0.2cm}
\begin{array}{c}
\textbf{U}_{1i} \\
\textbf{U}_{2i} \\
\textbf{U}_{3i}
\end{array}\hspace{-0.2cm}
\right)$,
the requirement $\Delta V(t,\xi^t)<0$ is equivalent to the following conditions that there exist $\textbf{U}_{1i}, \textbf{U}_{2i}, \textbf{U}_{3i}\in\mathbb{R}^{(N\hspace{-0.05cm}+\hspace{-0.05cm}1)\times (N\hspace{-0.05cm}+\hspace{-0.05cm}1)}$ $(i\in\mathcal{I})$ such that the linear matrix inequalities: $\forall (i,j)\in\mathcal{I}\times\mathcal{I}$,
\begin{align}
&\textbf{P}+\textbf{YH}+\textbf{H}^T\textbf{Y}^T<0, \nonumber \\
&\left(\hspace{-0.2cm}
\begin{array}{ccc}
-\tau^2\textbf{P}_i+\Gamma\textbf{F}_1 & \Gamma\textbf{F}_2 & 0 \\
\star & \Gamma & 0\\
\star & \star  & \textbf{P}_j\\
\end{array}\hspace{-0.2cm}
\right)+\left(\hspace{-0.2cm}
\begin{array}{c}
\textbf{U}_{1i} \\
\textbf{U}_{2i} \\
\textbf{U}_{3i}
\end{array}\hspace{-0.2cm}
\right)\left(\hspace{-0.2cm}
\begin{array}{ccc}
\textbf{B}_i \hspace{-0.2cm}& \textbf{C}_i \hspace{-0.2cm}& -\textbf{I}
\end{array}\hspace{-0.2cm}
\right)+\left(\hspace{-0.2cm}
\begin{array}{ccc}
\textbf{B}_i \hspace{-0.2cm}& \textbf{C}_i \hspace{-0.2cm}& -\textbf{I}
\end{array}\hspace{-0.2cm}
\right)^T\left(\hspace{-0.2cm}
\begin{array}{c}
\textbf{U}_{1i} \\
\textbf{U}_{2i} \\
\textbf{U}_{3i}
\end{array}\hspace{-0.2cm}
\right)^T<0, \nonumber \\
\label{eq:theorem2conditionsm}
&\left(\hspace{-0.2cm}
\begin{array}{ccc}
\textbf{U}_{1i}\textbf{B}_i+\textbf{B}_i^T\textbf{U}_{1i}^T-\tau^2\textbf{P}_i+\Gamma\textbf{F}_1 & \textbf{B}_i^T\textbf{U}_{2i}^T+ \textbf{U}_{1i}\textbf{C}_i +\Gamma\textbf{F}_2 & \textbf{B}_i^T\textbf{U}_{3i}^T-\textbf{U}_{1i}\\
\star & \textbf{C}_i^T\textbf{U}_{2i}^T+\textbf{U}_{2i}\textbf{C}_i+\Gamma & \textbf{C}_i^T\textbf{U}_{3i}^T-\textbf{U}_{2i}\\
\star & \star & \textbf{P}_j-\textbf{U}_{3i}^T-\textbf{U}_{3i}\\
\end{array} \hspace{-0.2cm}
\right)< 0.
\end{align}

Clearly, if it holds \eqref{eq:theorem2conditionsm}, we have $\Delta V(t,\xi^t)<0$. Following the proof of Theorem 1, GS-ADMM under arbitrary sequence is exponentially stable. The proof is complete. $\Box$
\emph{Proof of Corollary S1:} After the linearization $\overline{\textbf{B}}_i=\textbf{B}_i+\textbf{C}_i\textbf{G}_i$, the system does not contain the nonlinear function. Based on the proof of Theorem 1, we easily get the condition in Eq. \eqref{eq:controlplane} in the main body by removing the Eq. \eqref{eq:nonlinearfunctionineq} from the Eq. \eqref{eq:deltav23}.

]

\twocolumn[
Following the proof of Theorem 2, we set
\begin{align}
\textbf{y}=\left(\hspace{-0.2cm}
\begin{array}{c}
\overline{\xi}^{t} \\
\overline{\xi}^{t+1}
\end{array}\hspace{-0.2cm}
\right),\ \ \
\textbf{P}=\left(\hspace{-0.2cm}
\begin{array}{ccc}
-\tau^2\textbf{P}_i & 0 \\
\star & \textbf{P}_j\\
\end{array}\hspace{-0.2cm}
\right),\ \ \
\textbf{H}=\left(\hspace{-0.2cm}
\begin{array}{ccc}
\overline{\textbf{B}}_i \hspace{-0.2cm}& -\textbf{I}
\end{array}\hspace{-0.2cm}
\right), \ \ \ \textbf{Y}=\left(\hspace{-0.2cm}
\begin{array}{c}
\textbf{U}_{1i} \\
\textbf{U}_{3i}
\end{array}\hspace{-0.2cm}
\right). \nonumber
\end{align}
Thus, it easily holds the condition in Eq. \eqref{eq:theorem3condition-1} in the main body.  The proof is complete. $\Box$

$\\$
\textbf{\emph{Proof of Corollary 1:}} By employing common quadratic Lyapunov functions (CQLFs) \cite{Mason2004clf}, that is, $V(t,\xi^t)=(\xi^t)^T\textbf{P}\xi^t$ with $\textbf{P}=\textbf{P}^T>0$, similar to the proof of Theorem 1, Corollary 1 is easily proved. $\Box$

$\\$
\textbf{\emph{Proof of Theorem 3:}} Based on the Theorem 2, substituting $\widehat{\textbf{B}}_i=\textbf{B}_i$ by $\widehat{\textbf{B}}_i=\textbf{B}_i+\textbf{D}_i\textbf{K}_i$ in the conditions \eqref{eq:theorem2conditionsm} with $\textbf{U}_{1i}=0$ and $\textbf{U}_{2i}=0$ yields
\begin{align}
\label{eq:theorem3conditionsm}
\left(\hspace{-0.2cm}
\begin{array}{ccc}
-\tau^2\textbf{P}_i+\Gamma\textbf{F}_1 & \Gamma\textbf{F}_2 & \textbf{B}_i^T\textbf{U}_{3i}^T+(\mathbb{I}_{N+1}^i\textbf{V})^T\\
\star & \Gamma & \textbf{C}_i^T\textbf{U}_{3i}^T\\
\star & \star & \textbf{P}_j-\textbf{U}_{3i}^T-\textbf{U}_{3i}\\
\end{array} \hspace{-0.2cm}
\right)< 0.
\end{align}

When the condition in \eqref{eq:theorem3conditionsm} is held, the result then follows from Theorem 2, and $\textbf{P}_j-\textbf{U}_{3i}^T-\textbf{U}_{3i}< 0\Rightarrow \textbf{U}_{3i}^T+\textbf{U}_{3i}> \textbf{P}_j$. Since $\textbf{P}_j>0$, $\textbf{U}_{3i}^T+\textbf{U}_{3i}>0$ implying that $\textbf{U}_{3i}$
is full ranked. Let $\mathbb{I}_{N+1}^i\textbf{V}=\textbf{U}_{3i}\textbf{D}_i\textbf{K}_i$. Thus, $\textbf{D}_i\textbf{K}_i=\textbf{U}_{3i}^{-1}\mathbb{I}_{N+1}^i\textbf{V}\Rightarrow \textbf{K}_i=\frac{v_i}{d_i}(\textbf{U}_{3i}^{-1}\mathbb{I}_{N+1}^i)^T$. The proof is complete. $\Box$

$\\$
\textbf{\emph{Proof of Corollary 2:}} Similar to the proof of Theorem 3, we consider the Theorem 2 with $i=j=1$. Substituting $\widehat{\textbf{B}}=\textbf{B}$ by $\widehat{\textbf{B}}=\textbf{B}+\textbf{D}\textbf{K}$ in the conditions \eqref{eq:theorem2conditionsm} with $i=j=1$, $\textbf{U}_{1}=0$ and $\textbf{U}_{2}=0$ yields
\begin{align}
\label{eq:corollary2conditionsm}
\left(\hspace{-0.2cm}
\begin{array}{ccc}
-\tau^2\textbf{P}+\Gamma\textbf{F}_1 & \Gamma\textbf{F}_2 & \textbf{B}^T\textbf{U}_{3}^T+(\textbf{U}_{3}\textbf{D}\textbf{K})^T\\
\star & \Gamma & \textbf{C}^T\textbf{U}_{3}^T\\
\star & \star & \textbf{P}-\textbf{U}_{3}^T-\textbf{U}_{3}\\
\end{array} \hspace{-0.2cm}
\right)< 0.
\end{align}

When the condition \eqref{eq:corollary2conditionsm} is held, the result then follows from Theorem 2, and $\textbf{P}-\textbf{U}_{3}^T-\textbf{U}_{3}< 0\Rightarrow \textbf{U}_{3}^T+\textbf{U}_{3}> \textbf{P}$. Since $\textbf{P}>0$, $\textbf{U}_{3}^T+\textbf{U}_{3}>0$ implying that $\textbf{U}_{3}$
is full ranked. Because $\textbf{D}$ is a full rank matrix, $\textbf{V}$ satisfying $\textbf{U}_{3}\textbf{D}=\textbf{D}\textbf{V}$ are nonsingular and hence invertible. Then $\textbf{U}_{3}\textbf{D}\textbf{K}=\textbf{D}\textbf{V}\textbf{K}$. Let $\textbf{H}=\textbf{V}\textbf{K}$. Thus, $\textbf{U}_{3}\textbf{D}\textbf{K}=\textbf{D}\textbf{H}$ and $\textbf{K}=\textbf{V}^{-1}\textbf{H}$. The proof is complete. $\Box$

$\\$
\textbf{\emph{Proofs of Corollary C1 and Corollary C2:}} Similar to the proofs of Theorem 1, 2, and Corollary 1, Corollary C1 and Corollary C2 are easily proved. $\Box$

$\\$
\textbf{\emph{Proofs of Corollary C3:}} The proofs of Corollary C3 is similar to the proofs of Theorem 3. $\Box$

$\\$
\textbf{\emph{Proof of Corollary C4:}} Before proving Corollary C4, we first introduce the Schur complement Lemma.

\textbf{Lemma L1} (Schur complement) \cite{Boyd1994lmi} For a given symmetric matrix $\textbf{S}=\left(\hspace{-0.2cm}
\begin{array}{cc}
\textbf{S}_{11}&\textbf{S}_{12}\\
\star  & \textbf{S}_{22} \\
\end{array}\hspace{-0.2cm}
\right)$, where $\textbf{S}_{11}\in\mathbb{R}^{r\times r}$, the following conditions are equivalent:
\begin{itemize}
  \item 1) $\textbf{S}<0$;
  \item 2) $\textbf{S}_{11} < 0, \textbf{S}_{22} - \textbf{S}_{12}^T\textbf{S}_{11}^{-1}\textbf{S}_{12} < 0$;
  \item 3) $\textbf{S}_{22} < 0, \textbf{S}_{11} - \textbf{S}_{12}\textbf{S}_{22}^{-1}\textbf{S}_{12}^T < 0$.
\end{itemize}

Consider the CQLFs \cite{Mason2004clf} $V(t,\xi^t)=(\xi^t)^T\textbf{P}\xi^t$ with $\textbf{P}=\textbf{P}^T>0$, it holds
\begin{align}
\label{eq:corollary11condition1}
(\overline{\textbf{B}}_i+\textbf{D}_i\textbf{K}_i)^T\textbf{P}(\overline{\textbf{B}}_i+\textbf{D}_i\textbf{K}_i)-\tau^2\textbf{P} < 0, \ \ \forall i\in\mathcal{I}.
\end{align}
Generally, the problem of solving numerically \eqref{eq:corollary11condition1} for $(\textbf{P}, \textbf{K}_i)$ is very difficult since it is nonconvex. In order to make this problem numerically well tractable, applying the Schur complement formula  to \eqref{eq:corollary11condition1} yields
]

\twocolumn[
\begin{align}
\left(\hspace{-0.2cm}
\begin{array}{cc}
-\textbf{P}^{-1} &\overline{\textbf{B}}_i+\textbf{D}_i\textbf{K}_i\\
\star  & -\tau^2\textbf{P} \\
\end{array}\hspace{-0.2cm}
\right)< 0, \ \ \forall i\in\mathcal{I}. \nonumber
\end{align}
The above form is still nonlinear due to the occurrence of terms $\textbf{P}^{-1}$ and $\textbf{P}$. To overcome this problem, introduce the substitution $\textbf{Q}=\textbf{P}^{-1}$ and then multiply the result from the left and the right by $\mbox{diag}(\textbf{I}, \textbf{Q})$ to obtain
\begin{align}
\label{eq:corollary3condition}
\left(\hspace{-0.2cm}
\begin{array}{cc}
-\textbf{Q} &\overline{\textbf{B}}_i\textbf{Q}+\textbf{D}_i\textbf{K}_i\textbf{Q}\\
\star  & -\tau^2\textbf{Q} \\
\end{array}\hspace{-0.2cm}
\right)< 0, \ \ \forall i\in\mathcal{I}.
\end{align}
Let $\textbf{N}_i=\textbf{K}_i\textbf{Q}$. \eqref{eq:corollary3condition} is written as the condition in Eq. \eqref{eq:corollaryc4condition}. It is straightforward to see that \eqref{eq:corollary3condition} with $\textbf{N}_i=\textbf{K}_i\textbf{Q}$ is numerically solvable. Thus, $\textbf{K}_i=\textbf{N}_i\textbf{Q}^{-1}$. The proof is complete. $\Box$

$\\$
\textbf{\emph{Proofs of Corollary C5:}} The proofs of Corollary C5 is similar to the proofs of Corollary 2. $\Box$

$\\$
\textbf{\emph{Proofs of Corollary C6:}} The proofs of Corollary C6 is similar to the proofs of Corollary C4. $\Box$

\section{Numerical Experiments}
\label{sec:experiments}
In this section, three numerical experiments are used to demonstrate our theoretical results for the three challenges in the introduction. We will do three experiments to confirm the convergence of GS-ADMM and PJ-ADMM. The first experiment is used to verify the linear convergence of GS-ADMM under arbitrary sequence, The second and third experiments are used to verify the linear convergence of PJ-ADMM. In addition, designing parameter controller to make divergent PJ-ADMM convergent is shown in the counter example \cite{Chen2016ADMM} in the main body.

We consider the square of $\ell_2$ or $\ell_1$-minimization problems for finding regression or sparse solutions of a linear system:
\begin{align}
\label{eq:ell2problem}
\min_{\textbf{x}}\ \frac{1}{2}\|\textbf{x}\|^2,\ \ \ \emph{s.t.}\ \textbf{A}\textbf{x} =\textbf{c}, \\
\label{eq:ell1problem}
\min_{\textbf{x}}\ \|\textbf{x}\|_1,\ \ \ \emph{s.t.}\ \textbf{A}\textbf{x} =\textbf{c},
\end{align}
where $\textbf{x}\in\mathbb{R}^{m}$, $\textbf{A}\in\mathbb{R}^{m\times \overline{N}}$ and $\textbf{c}\in\mathbb{R}^{m}$. They have been widely used in signal and image processing, statistics, and machine learning. Suppose that the data are partitioned into $n$ blocks: $\textbf{x} =(\textbf{x}_1,\textbf{x}_2,\cdots,\textbf{x}_n)$ and $\textbf{A} = (\textbf{A}_1, \textbf{A}_2,\cdots,\textbf{A}_n)$, $\overline{N}=\sum_{i=1}^nn_i$, and $n_i$ is the dimensions of the variable $\textbf{x}_i$.

\begin{figure*}
\vskip -0.0in
\begin{center}
\centerline{\includegraphics[width=2\columnwidth]{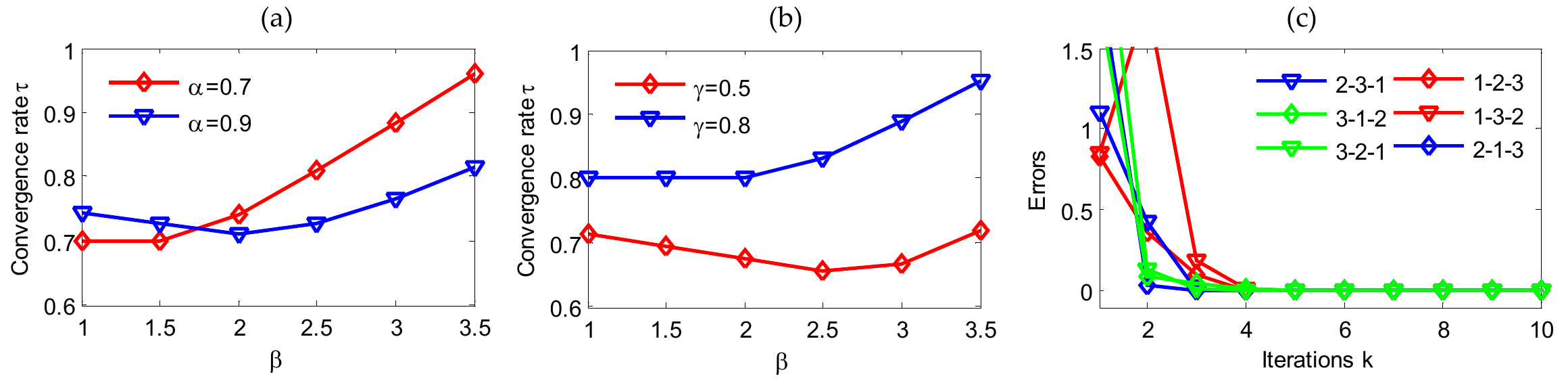}}
\vskip -0.05in
\caption{(a) plots the convergence rates with different $\beta$ and $\alpha$ when $\gamma=0.7$. (b) plots the convergence rates with different $\beta$ and $\gamma$ when $\alpha=0.8$. (c) also plots convergence curves of ADMM under arbitrary switching sequences with $\gamma=0.8$, $\beta=3$ and $\alpha=0.8$. }
\label{fig:example1}
\end{center}
\vskip -0.2in
\end{figure*}

\textbf{Experiment 1.} Consider the $\ell_2$ problem \eqref{eq:ell2problem} as the following strongly convex minimization with $x_1,x_2$ and $x_3$:
\begin{equation}
\min_{x_1,x_2,x_3}\ 0.1x_1^2+0.2x_2^2+0.1x_3^2,\ \ \ \emph{s.t.}\ \textbf{A}_1x_1+\textbf{A}_2x_2+\textbf{A}_3x_3 =\textbf{0},
\label{eq:strongconvexexample1}
\end{equation}
where $\textbf{A}_1=(0.1,-0.2,0.3)^T$, $\textbf{A}_2=(-0.3,-0.2,0.2)^T$ and $\textbf{A}_3=(0.1,-0.1,0.1)^T$.
Using the Proposition 1 in the main body and introducing Lagrange multipliers $\textbf{x}_4=(\nu_1,\nu_2,\nu_3)^T$, the example \eqref{eq:strongconvexexample1} is transformed into a linear switched system \eqref{eq:lineartransformgsdtsds} with the state $\textbf{x}=(x_1,x_2,x_3,\nu_1,\nu_2,\nu_3)^T$.
Using the condition \eqref{eq:corollaryc1condition-3} in our Corollary C1, the linearized GS-ADMM under arbitrary sequence on all the $x_1$, $x_2$ and $x_3$-subproblems will be convergent to solve the problem \eqref{eq:strongconvexexample1}. Given the fixing values $\gamma$, $\beta$, $\alpha=\alpha_1=\alpha_2=\alpha_3$ and $\tau$, the feasibility of the condition \eqref{eq:corollaryc1condition-3} is a semidefinite program with variables $\textbf{P}$, and they are easily solved by the LMI toolbox in the Matlab. We find the minimal rate $\tau$ by performing a binary search over $\tau$ such that the linear matrix inequality $\Phi\preceq 0$ is satisfied. The results are shown in Fig. \ref{fig:example1} (a) and (b) for a wide range of $\beta$, for several choices of $\alpha$ and $\gamma$. Moreover, the convergence curves in Fig. \ref{fig:example1} (c) show GS-ADMM under arbitrary sequence is linear convergent.

$\\$
In the next two experiments, we create real data for the convergence of PJ-ADMM. A solution $\textbf{x}^\star$ is randomly generated with $p$ $(p<n)$ nonzeros drawn from the standard Gaussian distribution. $\textbf{A}$ is also randomly generated from the standard Gaussian distribution, and its columns are normalized. $\textbf{x}$ and $\textbf{A}$ are partitioned evenly into $n$ blocks. The vector $c$ is then computed by $\textbf{c} = \textbf{A}\textbf{x}^\star + \delta$, where $\delta\sim N(0,\sigma^2\textbf{I})$ is Gaussian noise with standard deviation $\sigma$. We will find the minimal rate $\tau$ by performing a binary search over $\tau$ such that the linear matrix inequalities $\widehat{\Upsilon}\preceq 0$ or $\Upsilon\preceq 0$ is satisfied. (their feasibilities are a semidefinite program with variables $\textbf{P}$, and they are easily solved by the LMI toolbox in the Matlab.) In addition, we also measure the relative error $\frac{\|\textbf{x}^k-\textbf{x}^\star\|_2}{\|\textbf{x}^\star\|_2}$ in the parallel ADMM.

]

\twocolumn[

\begin{figure*}
\vskip -0.0in
\begin{center}
\centerline{\includegraphics[width=2\columnwidth]{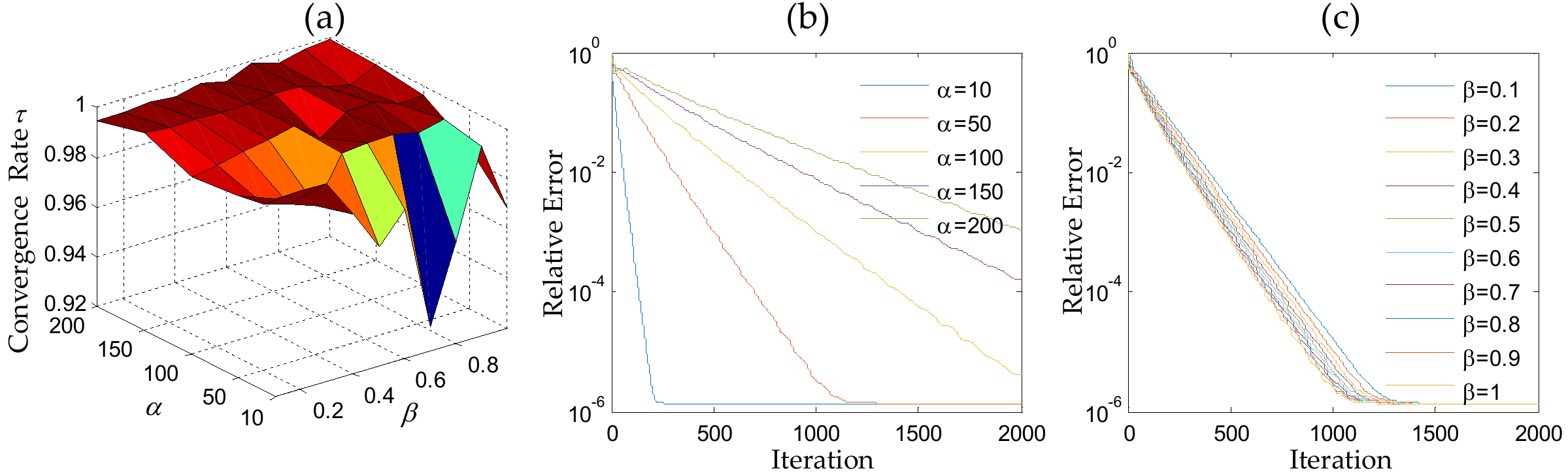}}
\vskip -0.15in
\caption{(a) plots the convergence rates $\tau$ with different $\beta$ and $\alpha$ when $\gamma=1$ and $n=10$. (b) plots the recover error with different $\alpha$ when $\beta=0.7$ and $\gamma=1$. (c) also plots the recover error with different $\beta$ when $\alpha=50$ and $\gamma=1$. }
\label{fig:experiment1}
\end{center}
\vskip -0.2in
\end{figure*}

\textbf{Experiment 2.} Consider $\overline{N} = 1000$, $m = 2000$, $p = 1000$ and the standard deviation of noise $\sigma$ is set to be $10^{-6}$, respectively. We set the number of blocks $n=10$. Since the activation function is $g(\textbf{x})=\textbf{x}=\nabla\frac{1}{2}\|\textbf{x}\|^2$ in the problem \eqref{eq:ell2problem}, it can be linearly transformed into the linear system \eqref{eq:lineartransformpjdtsds}. Using the Corollary C2, we search the convergence rates $\tau$ by solving the LMIs condition \eqref{eq:corollaryc2condition-2}. when $\gamma=1$, $\beta$ changes from $0.1$ to $1$, and five values of $\alpha=\alpha_i$ $(1\leq i\leq 10)$ are spaced between $10$ and $200$. The minimal convergence rates $\tau$ are shown in Fig. \ref{fig:experiment1} (a). For example, when $\alpha=10$ and $\beta=0.7$, we obtain $\tau=0.9294$ and $\tau^{180}=1.8901\times 10^{-6}$, that is, PJ-ADMM will convergent, and Fig. \ref{fig:experiment1} (b) also verifies the blue curve downs to $10^{-6}$ after the $k=180$ iterations. Moreover, the convergence curves in Fig. \ref{fig:experiment1} (b) and (c) show PJ-ADMM is convergent.

\begin{figure*}
\vskip -0.0in
\begin{center}
\centerline{\includegraphics[width=2\columnwidth]{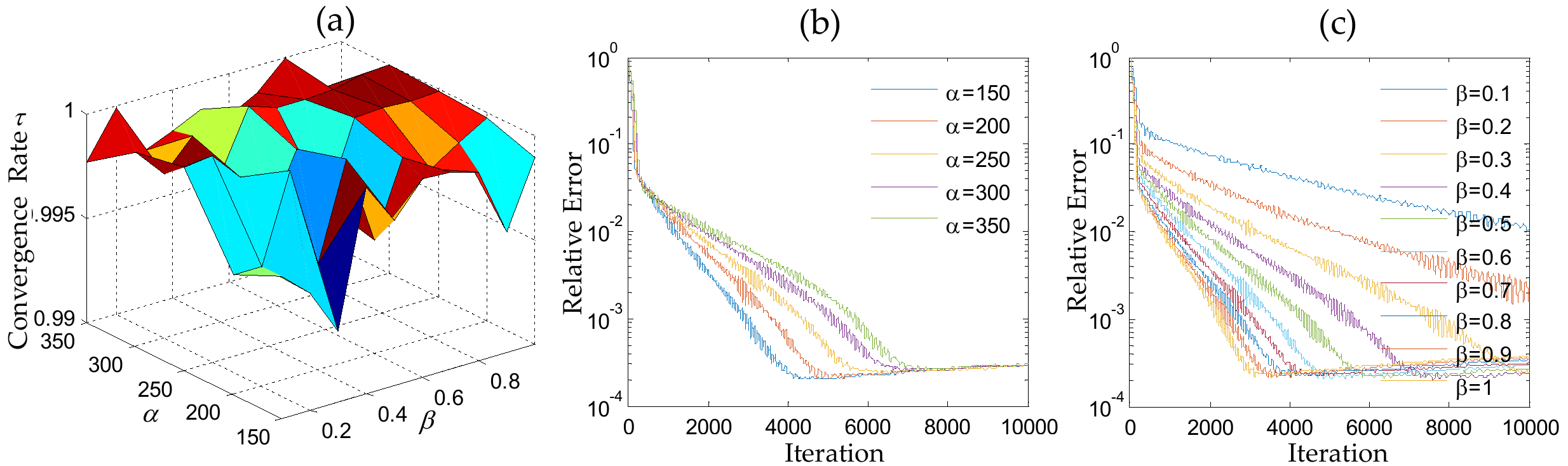}}
\vskip -0.15in
\caption{(a) plots the convergence rates $\tau$ with different $\beta$ and $\alpha$ when $\gamma=1$ and $n=20$. (b) plots the recover error with different $\alpha$ when $\beta=0.6$ and $\gamma=1$. (c) also plots the recover error with different $\beta$ when $\alpha=200$ and $\gamma=1$. }
\label{fig:experiment2}
\end{center}
\vskip -0.25in
\end{figure*}

$\\$
\textbf{Experiment 3.} Consider $\overline{N} = 2000$, $m = 1000$, $p = 100$ and the standard deviation of noise $\sigma$ is set to be $10^{-4}$, respectively. We set the number of blocks $n=20$. The shrinkage function $g(\textbf{x})=\max\{|\textbf{x}|-r,0\}\hbox{sign}(\textbf{x})$ can be regarded as an activation function since it is usually used to solve the $\ell_1$-problem \eqref{eq:ell1problem}. Moreover, $g(\textbf{x})$ satisfies $g(\textbf{x})=\rho\textbf{x}$, where $0\leq\rho<1$ and $\rho=\left\{
\begin{array}{ll}
0, & \hbox{$|\textbf{x}|\leq r$}, \\
(|\textbf{x}|-r)/|\textbf{x}|, & \hbox{$|\textbf{x}|>r$}.\\
\end{array}
\right.$. Therefore, the problem \eqref{eq:ell1problem} can be linearly transformed into the linear system \eqref{eq:lineartransformpjdtsds}. Similar to Experiment 2, we obtain the minimal convergence rates $\tau$ shown in Fig. \ref{fig:experiment2} (a) by using the LMIs condition \eqref{eq:corollaryc2condition-2} in the Corollary C2. Moreover, Fig. \ref{fig:experiment2} (b) and (c) plot the convergence curves to show that PJ-ADMM is convergent. We observe that the iterations will decrease with $\beta$ from $0.1$ to $1$ in Fig. \ref{fig:experiment2} (c). However, the convergence rate $\tau$ does not decrease. The plausible reason is the linearization of the shrinkage function. Fortunately, $\tau<1$ can ensure the convergence of PJ-ADMM.





]

\end{document}